\documentclass[a4paper,aps,prb,twocolumn,floatfix,showpacs,superscriptaddress,footnoteinbib]{revtex4-1}
\usepackage{graphics}
\usepackage{epsfig}
\usepackage{bbm}
\usepackage{times}
\usepackage{xcolor}   
\usepackage{amsfonts}
\usepackage{amsmath}
\usepackage{amssymb}
\usepackage{amsthm}
\usepackage{hyperref} 
\usepackage{wasysym}
\usepackage{bm} 

\newcommand{\overbar}[1]{\mkern 1.5mu\overline{\mkern-1.5mu#1\mkern-1.5mu}\mkern 1.5mu}

\begin{document}

\title{
Spin-polarized voltage probes for helical edge state: a model study
}

\author{Vivekananda Adak}
\affiliation{Department of Physical Sciences, IISER Kolkata, Mohanpur, West Bengal 741246, India.}
\author{Krishanu Roychowdhury}
\affiliation{Department of Physics, Stockholm University, SE-106 91 Stockholm, Sweden.}
\author{Sourin Das}
\affiliation{Department of Physical Sciences, IISER Kolkata, Mohanpur, West Bengal 741246, India.}
\email{vivekanandaaadak@gmail.com, krishanu.1987@gmail.com, sdas@physics.du.ac.in}

\begin{abstract}
Theoretical models of a spin-polarized voltage probe (SPVP) tunnel-coupled to the helical edge states (HES) of a quantum spin Hall system (QSHS) are studied. Our first model of the SPVP comprises $N_{P}$ spin-polarized modes (subprobes), each of which is locally tunnel-coupled to the HES, while the SPVP, as a whole, is subjected to a self-consistency condition ensuring zero average current on the probe. We carry out a numerical analysis which shows that the optimal situation for reading off spin-resolved voltage from the HES depends on the interplay of the probe-edge tunnel-coupling and the number of modes in the probe ($N_P$). We further investigate the stability of our findings by introducing Gaussian fluctuations in {\it{(i)}} the tunnel-coupling between the subprobes and the HES about a chosen average value and {\it{(ii)}} spin-polarization of the subprobes about a chosen direction of the net polarization of SPVP. We also perform a numerical analysis corresponding to the situation where four such SPVPs are implemented in a self-consistent fashion across a ferromagnetic barrier on the HES and demonstrate that this model facilitates the measurements of spin-resolved four-probe voltage drops across the ferromagnetic barrier.  As a second model, we employ the edge state of a quantum anomalous Hall state (QAHS) as the SPVP which is tunnel-coupled over an extended region with the HES. A two-dimensional lattice simulation for the quantum transport of the proposed device setup comprising a junction of QSHS and QAHS is considered and a feasibility study of using the edge of the QAHS as an efficient spin-polarized voltage probe is carried out including disorder. 
\end{abstract}
\maketitle
%
\section{Introduction}
Transport of electron spin with minimal loss of polarization and coherence over networks which are controllable by all electrical means is highly desirable for spintronics~\cite{Igor2004Spin, Atsufumi2020} and quantum information applications~\cite{Awschalom1988}. In particular, the surface states of two-dimensional and three-dimensional topological insulators, which possess spin-momentum locked spectra, can act as a resource for such applications~\cite{Pesin2012Spin, He2019topology, Yongbing2015}. Helical edge states (HES)~\cite{Wu2006, Xu2006, Maciejko2009} of quantum spin Hall state (QSHS)~\cite{kane2005quantum, kane2005z, bernevig2006quantum, bernevig2006quantumnature, Konig2007, Liu2008, Roth2009}, which are one-dimensional modes lying on the edges carrying electrons with their spin locked to their momenta ({\it {\it e.g.}}, right movers being spin-up and left movers being spin-down along a chosen spin-quantization axis), is one such state that constitutes the main topic of discussion in this manuscript. If we intend to exploit these states for spintronics applications, it requires to devise a way for carrying our spin-resolved measurements on the edge. Experimental attempts of probing these states are already in place, for example, a less invasive detection of these edge states was carried out by using mesoscale SQUID loop in Ref.~\onlinecite{Nowack2013} whereas a more invasive one involving the injection of electrical current into the edge was carried out in Ref.~\onlinecite{brune2012spin}. In this manuscript, we carry forward the latter idea and explore the possibility of using a spin-polarized voltage probe for reading off the local spin-resolved voltages on a helical edge. In particular, we formulate the problem to address a situation which is analogous to the six-probe Hall bar setup (two current probes and four voltage probes, as in Fig.~\ref{fig:figg4}) of Ref.~\onlinecite{Ferry2015}, involving a  quantum point contact (QPC) which is routinely used in measurements of Hall voltages. It should be noted that the idea of spin-polarized injection to HES using a ferromagnetic electrode has been theoretically explored in Ref.~\onlinecite{arrachea2011chiral}. In fact, an earlier  experimental realization engaging a Hall bar type setup can be found in Ref.~\onlinecite{valenzuela2006direct}.   

\begin{figure}
\centering
\includegraphics[width=1.0\columnwidth]{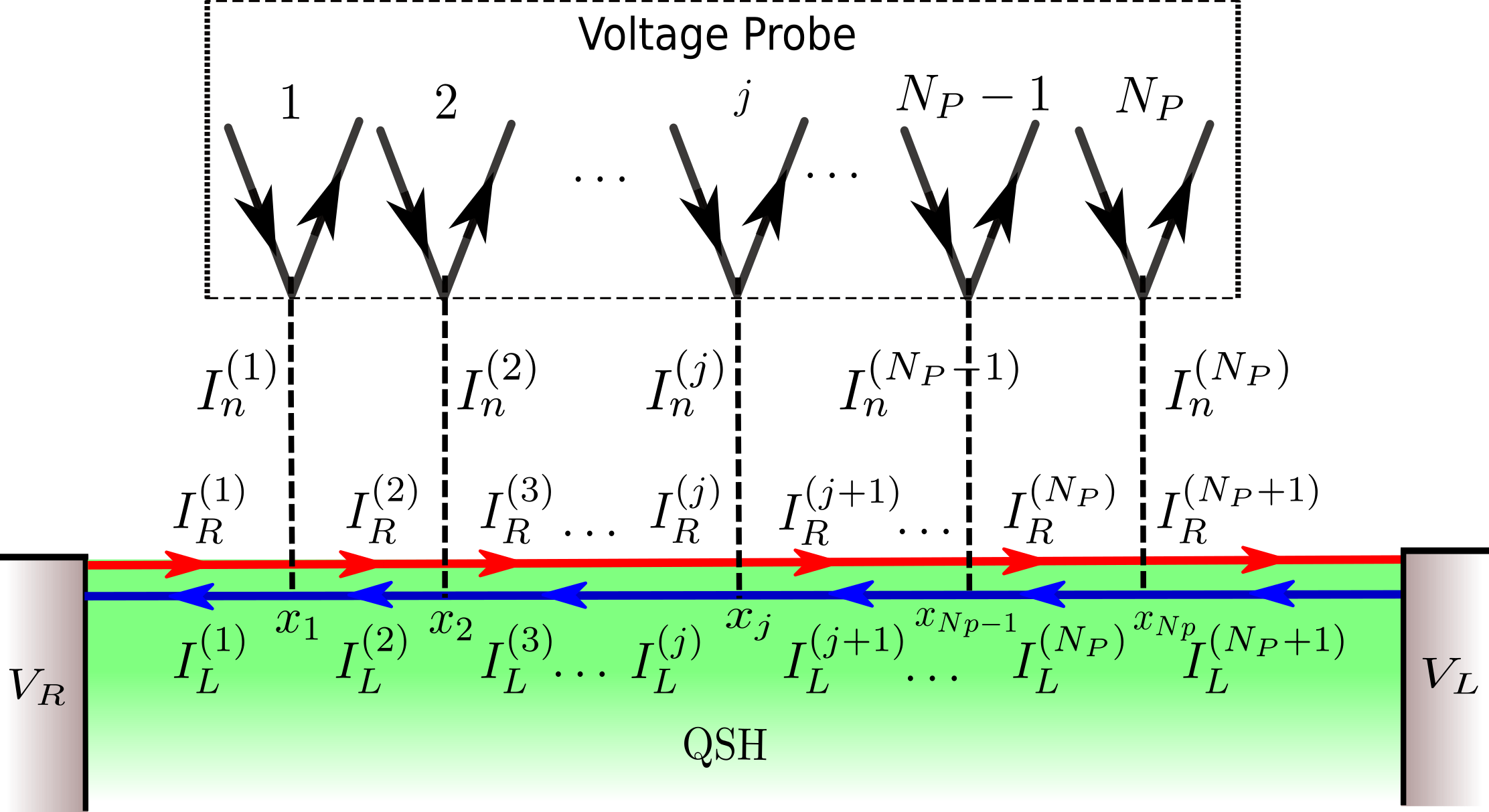}
\caption{Schematic of the setup involving a SPVP, which is modelled as a collection of chiral, spin-polarized, one-dimensional modes, tunnel-coupled to the helical edge state subjected at a finite bias voltage ($V_R-V_L$). The red (blue) line represents the spin-up (down) edge. The shaded region in green represents the QSHS and the two gray patches on the two sides represents the leads across which the voltage bias is applied.}
\label{fig:figg1}
\end{figure}

We consider a situation where a single backscatterer is placed on the HES which is carrying a net current due to the electrical bias applied across the edge state. We then place four spin-polarized voltage probes, two on the right side, and two on the left side of the backscatterer. The polarization directions of the probes are such chosen that the pair of probes on each side of the backscatterer has one fully polarized along the $z$-axis (call it an ``up-polarized") while the other fully polarized along the $-z$-axis (call it a ``down-polarized"). We expect the up-polarized probe would couple primarily to the right movers (spin-up electrons on the edge) and the down-polarized probe will couple mostly to the left movers (spin-down electrons on the edge) as we have assumed spin conserving electron tunneling between the probes and the helical edge. We will present a detailed analysis which explores the possibility of such spin-resolved coupling~\cite{das2011spin} of the $z$-polarized voltage probes. Note that a number of possible sources of backscattering on a HES have been studied, however, this work focuses primarily on the ballistic limit~\cite{Maciejko2009,Tanaka2011,Maciejko2012,Lezmy2012,Budich2012,Schmidt2012, Lunde2012,Eriksson2012,Jukka2013,Del2013,Eriksson2013-1,Eriksson2013-2, Altshuler2013,Geissler2014,Kainaris2014,Jukka2014,Pikulin2014,Dolcetto2016, Kimme2016, yuval2018}. 

For a pictorial representation of the situation described above, see Fig.~\ref{fig:figg4}. It is straightforward to note the analogy between the situation described in Fig.~\ref{fig:figg4}, (a) and Fig.~\ref{fig:figg4}, (c)  if one identifies the left (right) movers on the top and the bottom edge in the Hall setup with the left (right) movers of the HES. With such a setup, one can measure spin-resolved voltage drops (voltage drop between a pair of up or down probes) across the impurity which is nothing but the drop in voltages in the right and the left moving channels of the HES across the backscatterer. These voltages will be analogous to the longitudinal voltage drop ($V_{lo}$) in the Hall bar geometry measured across the QPC. We can also measure the voltage difference between the right and the left mover on the same side by using the up and the down-polarized probe on the same side of backscatterer and that will be analogous to the measurement of the Hall voltage ($V_{H}$) in the Hall bar geometry which should be a robust quantity, {\it i.e.}, independent of the strength of the backscatterer provided the voltage probes are ideal~\cite{MacDonald1973Edge, buttiker1988coherent, Yoseph2002, forster2007voltage, jacquet2012temperature}.

In this manuscript, we will demonstrate via numerical calculations that our model for the SPVP, when optimized appropriately, can lead to measured values of $V_{lo}$ and $V_{H}$ which indeed correspond to the spin-resolved voltage drops across the backscatterer with the latter an analog of Hall voltage on the HES. This, in turn, implies that our theoretical model for the spin-polarized probe is capable of measuring the spin-resolved voltages successfully and hence, could provide useful guidance to future experiments that are tuned to such objectives. 

The rest of the paper is organized as follows: In section~\ref{sectwo}, we lay the concept of spin-resolved voltage measurements on a HES tunnel-coupled to a $N_P$-subprobe SPVP and discuss the stability of the measurements in presence of Gaussian disorder in {\it{(i)}} the tunnel-coupling between the subprobes of the SPVP and the helical edge and {\it{(ii)}} the spin-polarization of the subprobes about a chosen direction. In section~\ref{secthree}, we demonstrate the six-probe setup discussed above and measure Hall response in presence of a backscatterer including Gaussian disorder in the individual SPVPs. Finally in section~\ref{secfour}, we simulate a device setup using the KWANT package~\cite{groth2014kwant} which comprises a quantum anomalous Hall state (QAHS) acting as the SPVP for the HES formed at the edge of the QSHS and also present a feasibility study. We summarize the results and conclude in section~\ref{secfive}.

\section{Spin-resolved voltage measurement on helical edge states}
\label{sectwo}

As already discussed in the introduction, our setup to measure the spin-resolved voltage on the HES of a QSHS engages an extended SPVP tunnel-coupled to the HES. The SPVP consists of multiple modes each of which supports spin injection into the HES as shown in Fig.~\ref{fig:figg1}. We also assume that the QSHS  is hosted on the $x$-$y$ plane with the relevant HES of the QSHS laying along the $x$-axis and described by the Hamiltonian
\begin{equation}
 \mathcal{H}_{\rm HES} = -\imath \hbar v_F \int_{-\infty}^{\infty} dx ~(\psi_R^\dagger \partial_x \psi_R - \psi_L^\dagger \partial_x \psi_L),
 \label{Hedge}
\end{equation}
where $v_F$ is the Fermi velocity, the operators $\psi_R^\dagger$ and $\psi_L^\dagger$ create electrons respectively in the right ($R$) and the left ($L$) propagating edge states with spinors $|n_R \rangle=[1~0]^T$ and $|n_L \rangle=[0~1]^T$ respectively implying the spin-polarization of the HES being along the $z$-axis, perpendicular to the plane hosting the QSHS as found in experiments~\cite{brune2012spin}. 

We model the SPVP as a collection of one dimensional modes with linear spectrum each of which is, henceforth, referred to as a subprobe. Each of these subprobes is regarded as a right moving chiral mode ($R'$) with spin-polarization given by the spinor $|n_{R'} \rangle \equiv [\cos(\theta/2+\pi/4)~\sin(\theta/2+\pi/4)e^{i\phi}]^T$ ($\theta,\phi$ are the polar and the azimuthal angle respectively of $|n_{R'} \rangle$ on the Bloch sphere). As the tunneling between the subprobes and the HES is taken to be local, hence, the chirality of the subprobes (being right or left movers) is of no consequence as far as tunneling current is concerned. The linear spectrum of the subprobes ensures that the tunneling current, in the weak tunneling limit, does not develop an energy dependence due to the variation in the density of states of the probe at the Fermi level as desirable for an ideal probe. Also the modelling of the voltage probe as a collection of subprobes is motivated from the fact that this proves a minimal model for the probe accommodating a large number of electronic degrees of freedom. Hence, the Hamiltonian for a subprobe is given by 
\begin{equation}
 \mathcal{H}_{\rm subprobe}^{(j)} = -\imath \hbar v_F \int_{-\infty}^{\infty} dx~ \psi_{R'j}^\dagger \partial_x \psi_{R'j},
 \label{Htip}
\end{equation}
so that $ \mathcal{H}_{\rm {SPVP}} = \sum_{j=1}^{N_P} \mathcal{H}_{\rm {subprobe}}^{(j)}$. Note that an offset of $\pi/4$ is introduced in the expression of ${|n_{R'j} \rangle}$ merely to set the range of the polarization angle of the SPVP to be $\theta\in[-\pi/2,\pi/2]$, symmetric about $\theta=0$. The tunneling Hamiltonian for electrons between the SPVP with $N_P$ number of subprobes and the HES is taken to be $\mathcal{H}_T = \sum_{j=1}^{N_P} \mathcal{H}_{T}^{(j)}$ such that,
\begin{equation}
 \mathcal{H}_{T}^{(j)} = \sum_{\eta} t_{\eta R'}^{(j)} \psi_{\eta}^\dagger(x_j) \psi_{R'j}(0) + {\rm h.c.},
 \label{htun}
\end{equation}
where the tunnel-coupling between the HES and the subprobe is taken to be such that for each subprobe, the tunneling is happening at $x=0$ of the subprobe coordinates and at $x=x_j$ for the corresponding HES coordinates; $t_{\eta R'}^{(j)}$ is the tunneling strength between the right or the left moving states [$\eta\in(R,L)$] of the HES and the chiral edge $R'$ representing the $j$-th subprobe, further expressed as $t_{\eta R'}^{(j)}=t'^{(j)}\gamma_{\eta R'}^{(j)}$, $t'^{(j)}$ being real. Note that the form of the tunneling Hamiltonian preserves the spin rotation symmetry of the electron. The quantity $\gamma_{\eta R'}^{(j)}$ denotes the overlap between the spinors $|n_{\eta}\rangle$ and $|n_{R'j}\rangle$: $\gamma_{\eta R'}^{(j)}=\langle n_{\eta}|n_{R'j}\rangle$. For an extended SPVP with multiple subprobes, we consider the following cases - 
\begin{itemize}
\item[{\it{(i)}}] All the subprobes have identical polarization set by an angle $\theta$ and also the tunneling strength $t'^{(j)}$ are taken to be  uniform across the junctions (with a magnitude $t'$). We refer to this as the {\it uniform} case.
\item[{\it{(ii)}}]  The tunneling strength $t'^{(j)}$ is nonuniform across the junctions and is characterized by a Gaussian distribution with mean $\overbar{t'}$ and standard deviation $\sigma_{t'}$ while the polarization angle $\theta$ of the subprobes fluctuates with a mean $\overbar{\theta}$ and standard deviation $\sigma_{\theta}$. We refer to this as the {\it disordered} case. 
\end{itemize}

Electron transport across a given tunneling point between a subprobe and the HES at $x=x_j$ (see Fig.~\ref{fig:figg1}) can be quantified in terms of a scattering matrix ${\cal S}_j$ corresponding to the Hamiltonian $\mathcal{H}=\mathcal{H}_{\rm HES}+\mathcal{H}^{(j)}_{\rm subprobe}+\mathcal{H}_{T}^{(j)}$. The wavefunctions associated with the incoming and the outgoing electrons at the tunneling point at $x=x_j$ (denoted $\Psi_{\alpha j}^{\rm in}$ and $\Psi_{\alpha j}^{\rm out}$ respectively) are related by the elements of ${\cal S}_j$:
\begin{equation}
 \Psi^{{\rm out}}_{\alpha j} = \sum_{\beta} s^{(j)}_{\alpha\beta}~ \Psi^{{\rm in}}_{\beta j},
 \label{scat_mat}
\end{equation}
where $\alpha,\beta\in (R,L,R')$ and the corresponding currents obey
\begin{equation}
 I^{\rm out}_{\alpha j} = \sum_{\beta} |s^{(j)}_{\alpha\beta}|^2~I^{\rm in}_{\beta j}.
 \label{curr_mat}
\end{equation}
For the SPVP to act as an ideal voltage probe, the net current flowing through it must be zero {\it i.e}. $I^{\rm in}_{\rm SPVP}-I^{\rm out}_{\rm SPVP}=0$ is the {\it voltage probe condition}, where $I^{\rm in}_{\rm SPVP}=\sum_{j=1}^{N_{p}} I^{\rm in}_{R'j}$ and $I^{\rm out}_{\rm SPVP}=\sum_{j=1}^{N_{p}} I^{\rm out}_{R'j}$. For an SPVP consisting of one subprobe only ($N_P=1$, tunneling strength $t'$, polarization $\theta$), this condition yields a probe voltage given by
\begin{align}
 V_{R'} = \frac{|s_{R'R}|^2V_R+|s_{R'L}|^2V_L}{|s_{R'R}|^2+|s_{R'L}|^2},
 \label{probvolt1}
\end{align}
when the HES is connected to a right and a left reservoir maintained at the voltages $V_R$ and $V_L$ respectively. In presence of a finite bias ($V_R-V_L\neq 0$), the HES develops a net magnetization along the $z$-axis, hence, in the weak tunneling limit ($t'\ll \hbar v_F$), the voltage measured by the SPVP is expected to be a sum of the average voltage $V_{\rm av}=(V_R+V_L)/2$ and an additional contribution which can be attributed to tunnel magnetoresistance~\cite{Gerstner2007Nobel}. Thus, we get
\begin{align}
 V_{R'}\vert_{{N_P}=1}=\frac{1}{2}(V_R+V_L) - \frac{1}{2}(V_R-V_L)\sin\theta. 
 \label{probvolt2}
\end{align}
Note that this expression is independent of the azimuthal angle $\phi$ as expected and the $\sin\theta$ dependence of the magnetoresistance instead of the standard $\cos\theta$ is due to our phase shifted definition of $\theta$. This expression of $ V_{R'}$ is obtained by using the explicit forms of the scattering matrix elements (see Appendix~\ref{appA} for details). In fact, the above form of $V_{R'}$ is valid for all values of $t'$  owing to the fact that the ratios $|s_{R'R}|^2/(|s_{R'R}|^2+|s_{R'L}|^2)=\cos^2(\theta/2)$ and $|s_{R'L}|^2/(|s_{R'R}|^2+|s_{R'L}|^2)=\sin^2(\theta/2)$ are independent of $t'$ though $t'$ induces spin flip scattering in the HES via higher order processes when $\theta \neq \pm \pi/2$. This leads to the interesting fact that the form of the magnetoresistance contribution, which is expected to exist in the weak tunneling limit, actually remains intact even at intermediate and strong tunneling limits. This magnetoresistance contribution in $V_{R'}$ is the key element for obtaining spin-resolved information of the HES by using the SPVP and it is the primary focus of this study. Evidently, when $\theta=-\pi/2$, the spinor $|n_{R'}\rangle$ aligns with $|n_R\rangle$ and the SPVP measures the voltage $V_{R'}=V_R$, while for $\theta=\pi/2$, $|n_{R'}\rangle$ aligns with $|n_L\rangle$ yielding a voltage measurement $V_{R'}=V_L$. However, these results holds only in the limit of few subprobes which we will discuss in detail later. For the case of multiple subprobes ($N_P>1$), the voltage probe condition implies
\begin{align}
 \sum_{j=1}^{N_P} (I^{\rm out}_{R'j}-e^2 V_{R'}/h) \equiv \sum_{j=1}^{N_P} I^{(j)}_{n}=0,
 \label{probvolt3}
\end{align}
where $I^{(j)}_{n}$ indicates the net current in the $j$-th subprobe. Also note that, as we have assumed all subprobes to be in equilibrium with a large reservoir, which is maintained at a voltage $V_{R'}$, hence, the incoming current in each of the subprobes is given by the Hall relation, $I^{\rm in}_{R'j}=e^2 V_{R'}/h, \forall j$ and the information regarding the values of $V_L, V_R$ enters Eq.~\ref{probvolt3} via $I^{\rm out}_{R'j}$ due to Eq.~\ref{curr_mat}. In general, to obtain an analytic expression for $V_{R'}$ by solving Eq.~\ref{probvolt3} is fairly complicated for a given set of values for $N_P$, $V_L$, and $V_R$. Hence, we solve Eq.~\ref{probvolt3} numerically and also set  $e^2/h=1$ henceforth, unless stated otherwise. 

\subsection{Transfer matrix approach for numerical analysis}

\begin{figure}
\centering
\includegraphics[width=1.0\columnwidth]{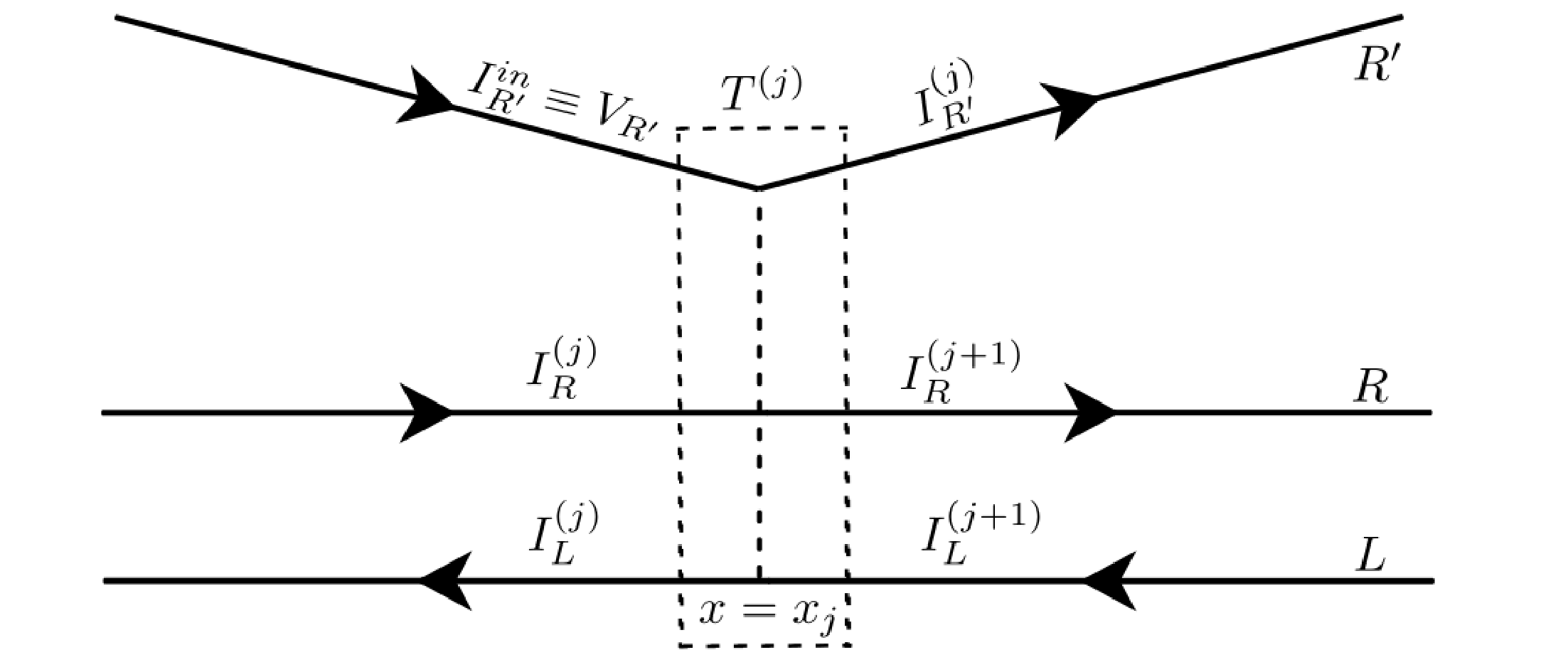}
\caption{Zoomed-in pictorial representation of a subprobe tunnel-coupled to the HES where $T$ represents the transfer matrix which connects the wavefunctions on the right side of the junction to that on its left side.}
\label{fig:figgxxx}
\end{figure}

To obtain $V_{R'}$ self-consistently for a given set of values for $V_R$, $V_L$, and $t'$ numerically, it is efficient to use a transfer matrix method which connects the left and right going currents on one side of the tunnel junction between the $j$-th subprobe and the HES at $x=x_j$ to that on the other side as shown in Fig.~\ref{fig:figgxxx}. To proceed further, we compartmentalize the HES into $N_P+1$ segments where the $j$-th segment is defined as the segment of the HES lying between the tunnel junction at $x=x_{j-1}$ and $x=x_{j}$ (see Fig.~\ref{fig:figg1}). Note that the $1$-st and the  $N_P+1$-th segment are connected only to one tunnel junction being the first and the last one. Also the incoming current in each subprobe (in units of $e^2/h$) is $V_{R'}$ irrespective of the position of the subprobe. Now the relation between currents on different segments can be written as:
\begin{align}
 I_R^{(j+1)} &= T^{(j)}_{RR} I_R^{(j)} + T^{(j)}_{RL}I_L^{(j)} + T^{(j)}_{RR'}V_{R'} \nonumber \\
 I_L^{(j+1)} &= T^{(j)}_{LR} I_R^{(j)} + T^{(j)}_{LL}I_L^{(j)} + T^{(j)}_{LR'}V_{R'} \nonumber \\
 I_{R'}^{(j)} &= T^{(j)}_{R'R} I_R^{(j)} + T^{(j)}_{R'L}I_L^{(j)} + T^{(j)}_{R'R'}V_{R'},
 \label{curr_mat3}
\end{align}
where $T^{(j)}_{\eta\eta'}$ denotes the transfer matrix elements at the $j$-th junction for the currents while $I_{\eta}^{(j)}$ ($\eta \in R,L$) corresponds to the right moving or the left moving current on the $j$-th segment. In this notation, the net current in the $j$-th subprobe is given by $I_{n}^{(j)}=I_{R'}^{(j)}-V_{R'}$. The elements ${T}_{\eta\eta'}^{(j)}$ can be expressed in terms of the elements of the scattering matrix,  ${\cal S}^{(j)}$,  and is given by 
\begin{align}
 T^{(j)}_{RR} &= |s^{(j)}_{RR}|^2 - |s^{(j)}_{RL}|^2|s^{(j)}_{LR}|^2/|s^{(j)}_{LL}|^2 \nonumber \\
 T^{(j)}_{RL} &= |s^{(j)}_{RL}|^2/|s^{(j)}_{LL}|^2 \nonumber \\
 T^{(j)}_{RR'} &= |s^{(j)}_{RR'}|^2 - |s^{(j)}_{RL}|^2|s^{(j)}_{LR'}|^2/|s^{(j)}_{LL}|^2 \nonumber \\
 T^{(j)}_{LR} &= -|s^{(j)}_{LR}|^2/|s^{(j)}_{LL}|^2 \nonumber \\
 T^{(j)}_{LL} &= 1/|s^{(j)}_{LL}|^2 \nonumber \\
 T^{(j)}_{LR'} &= -|s^{(j)}_{LR'}|^2/|s^{(j)}_{LL}|^2 \nonumber \\
 T^{(j)}_{R'R} &= |s^{(j)}_{R'R}|^2 - |s^{(j)}_{R'L}|^2|s^{(j)}_{LR}|^2/|s^{(j)}_{LL}|^2 \nonumber \\
 T^{(j)}_{R'L} &= |s^{(j)}_{R'L}|^2/|s^{(j)}_{LL}|^2 \nonumber \\
 T^{(j)}_{R'R'} &= |s^{(j)}_{R'R'}|^2 - |s^{(j)}_{R'L}|^2|s^{(j)}_{LR'}|^2/|s^{(j)}_{LL}|^2,
\end{align}
 and the expressions of the elements $s^{(j)}_{\eta\eta'}$ in terms of $t'$ and $\theta$ are given in Appendix~\ref{appA}. Also, we assume that there is no quantum coherence between the scattering of electrons at successive junctions between the HES and the subprobes. To obtain $V_{R'}$, the equations in Eq.~\ref{curr_mat3} are solved recursively for $j\in \{1,\dots,N_P\}$ subject to the voltage probe condition in Eq.~\ref{probvolt3} and boundary conditions given by $I_L^{N_P+1}=V_L$ and $I_R^{1}=V_R$. With this formulation discussed above, we can now study the cases mentioned previously, {\it i.e.}, the {\it uniform} case and the {\it disordered} case. 

\begin{figure}
\centering
\includegraphics[width=1.0\columnwidth]{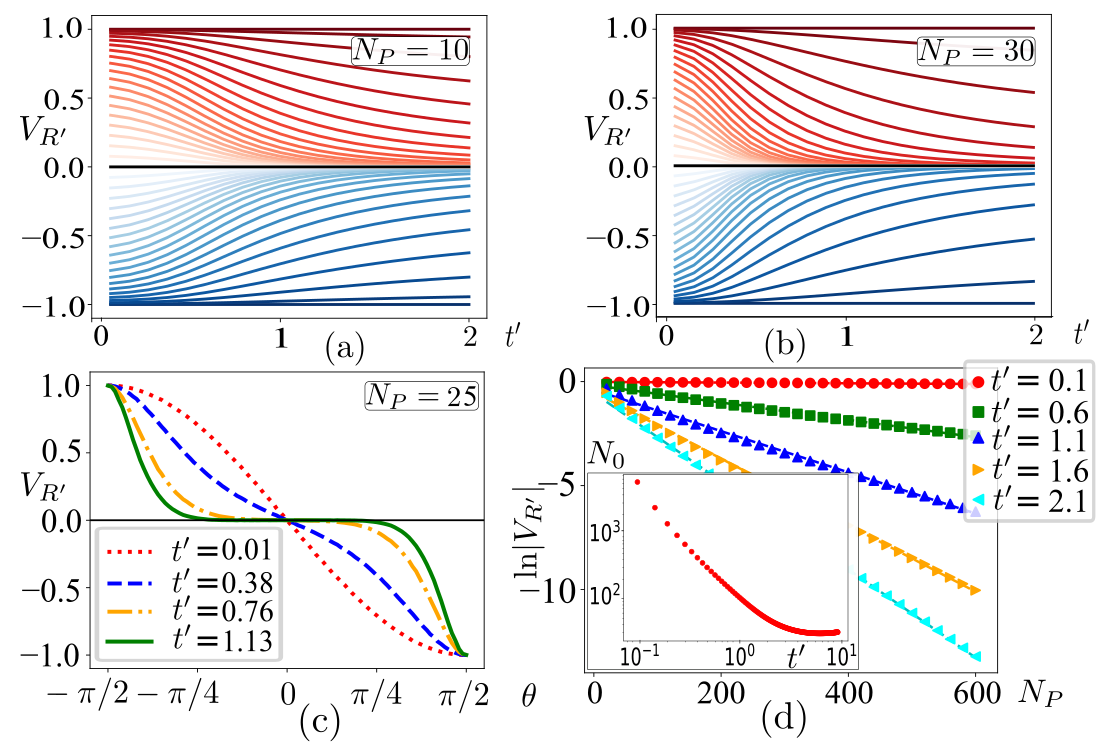}
\caption{Plot of $V_{R'}$ against tunneling strength $t'$ for a uniform SPVP with $N_P=10$ in (a) and $N_P=30$ in (b), where the HES is maintained at a voltage bias given by $V_R=-V_L=1$. Different curves correspond to different values of $\theta$ starting from $\theta=-\pi/2$ (the top most plot) to $\theta=\pi/2$ (the bottom most plot) in steps of $\pi/40$ (the offset along the $t'$-axis of these two figures is $t'=0.04$). (c) Plot of $V_{R'}$ as a function of $\theta$ at different values of $t'$ for $N_P=25$. 
(d) Plot of $V_{R'}$ as a function of number of subprobes $N_P$ in the SPVP at different values of $t'$ for $\theta=-1.37$. The inset shows the scale of this exponential decay, denoted $N_0$, plotted as a function of $t'$. The offset along the $N_P$-axis for this plot is $N_P=20$.}
\label{fig:figg2}
\end{figure}

\subsection{Uniform case}
The results for the uniform case with $N_P=10$ and $N_P=30$ subprobes are displayed in Fig.~\ref{fig:figg2} (a) and (b) respectively for  $V_R=-V_L=1$. These values for $V_R$, $V_L$ are taken for the numerical calculations as it corresponds to a zero $V_{\rm av}$ hence leaving behind a neat magnetoresistance contribution in Eq.~\ref{probvolt2}. From the plots in Fig.~\ref{fig:figg2} we note that the value of $V_{R'}$ decays  monotonically towards the average voltage $V_{\rm av}$ as we increase $t'$. This can be understood as follows. In the weak tunneling limit ($t'\ll \hbar v_F$), spin-flip scattering induced by the subprobes in the HES is minimal and hence $V_{R'}$ carries spin-resolved information, {\it i.e.}, it follows Eq.~\ref{probvolt2}. But as we increase the tunneling strength ($t'$), it leads to considerable spin-flip scattering in the HES at the junctions with the subprobes. The feedback between various subprobes amplifies the effect, hence, resulting in reduction of the magnetoresistance response appearing in Eq.~\ref{probvolt2} and eventually suppressing it completely in the large $t'$ limit. The rate at which $V_{R'}$ drifts towards $V_{\rm av}$ as a function of $t'$ depends on two factors: ({\it{i}}) the angle between the spin-polarization axis of the SPVP and the spin quantization axis of HES, {\it i.e.}, $\theta$, and ({\it{ii}}) the number of subprobes in the SPVP. Note that the case corresponding to $\theta=\pm \pi/2$ are pathological and in these two cases, $V_{R'}$ does not decay towards $V_{\rm av}$ as we increase $t'$ for the subprobes do not induce any spin-flip scattering in the HES and, hence, provides a perfect readout of the spin-resolved voltages.

\begin{figure}
\centering
\includegraphics[width=1.0\columnwidth]{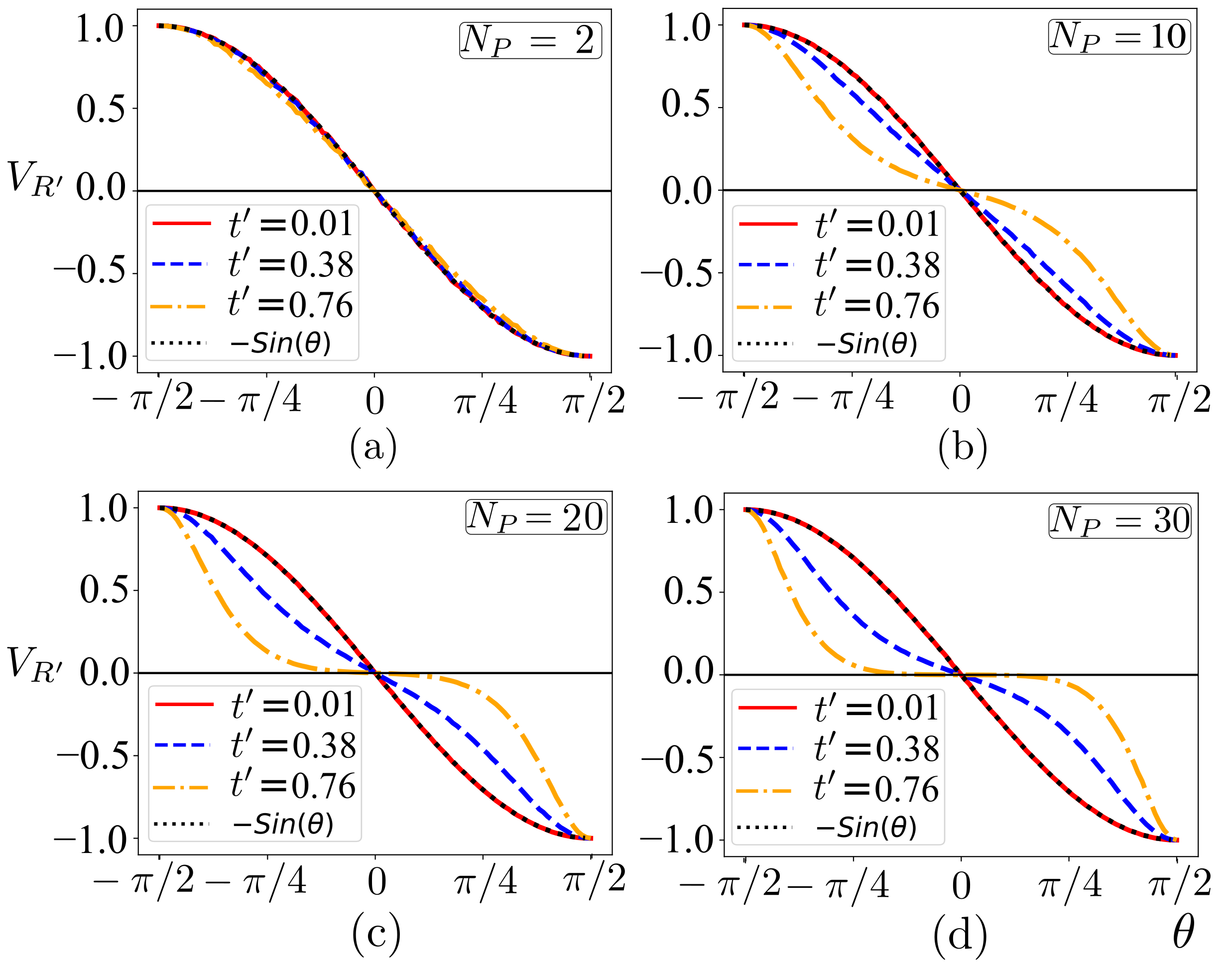}
\caption{Variation of $V_{R'}$ as a function of $\theta$ for $N_P=2$, $10$, $20$, and $30$ in (a), (b), (c), and (d) respectively.}
\label{fig:figg'}
\end{figure}

In Fig.~\ref{fig:figg2} (c) we have plotted $V_{R'}$ as a function of $\theta$ for different values of $t'$ and $N_{P}=25$. We note that the value of $V_{R'}$ reduces to $V_{\rm av}=0$ independent of $t'$ for $\theta=0$ owing to fact that the tunnel-coupling strengths of the SPVP with the spin-up (right mover) and the spin-down (left mover) channel are equal in this case (see the form of $t_{\eta R'}^{(j)}$ below Eq.~ \ref{htun}) hence, leading to a null magnetoresistance. We further note that larger values of $t'$ leads to stronger spin equilibration in the HES and hence, it forces $V_{R'}$ to stay close to $V_{\rm av}=0$ for most values of $\theta$ which are away from $\theta=\pm\pi/2$. But as we consider the value of $\theta$ close to $-\pi/2$ or $\pi/2$, the ratio between the tunneling strength for the spin-up and the spin-down channel with the SPVP becomes very large or small respectively, hence, leading to a strong suppression of spin equilibration in the HES. Consequently, $V_{R'}$ shows a large deviation from $V_{\rm av}$ implying that it retains spin-resolved information. Before we move on to Fig.~\ref{fig:figg2} (d), to elaborate further on the  $\theta$ dependence of $V_{R'}$, we will discuss the plot of $V_{R'}$ as a function of $\theta$ presented in Fig.~\ref{fig:figg'} for different value of $N_{P}$. We note that the plots for the case of $N_{P}=2$ for various values of $t'$ starting from a small value ($t'=0.01$) to a large one ($t'=0.76$) closely follows Eq.~\ref{htun}. This can be seen from the closeness of the plots for $V_{R'}$ to that of $\sin \theta$. So it is clear that the case of two subprobes does not immediately changes the scenario as far as deviation from Eq.~\ref{htun} is concerned. Only when we add multiple subprobes in the SPVP, it leads to a large deviation from Eq.~\ref{htun} away from $\theta=\pm\pi/2$ and we observe this very clearly in the cases such as $N_{P}=20$ and $N_{P}=30$ in Fig.~\ref{fig:figg'}.

In Fig.~\ref{fig:figg2} (d), we further demonstrate that for $\theta\neq \pm\pi/2$, $V_{R'}$ falls exponentially to $V_{\rm av}$ with increasing number of subprobes and, in the large $N_P$ limit, is given by 
\begin{align}
 V_{R'}=V_{\rm av} + [~V_{R'}\vert_{{N_P}=1} - V_{\rm av}\,]\,e^{-(N_P-1)/N_0}\,.
\label{exp}
\end{align}
This exponential decay is visible in the main panel of Fig.~\ref{fig:figg2} (d) for various values of $t'$. This indicates that the contribution to $V_{R'}$ due to the magnetoresistance effect gets exponentially suppressed with $N_P$; $N_0$ multiplied by the average spacing between the subprobes defines a characteristic length scale for the decay. The decay length, obtained from $N_0$, can be though of as a measure of the length scale over which the spin-up and the spin-down edge equilibrate leading to loss of spin-polarization of the HES induced by a voltage bias ($V_L$-$V_R$). The plot in Fig.~\ref{fig:figg2} (d) corresponds to a value of $\theta$ which is shifted from $-\pi/2$ by $0.2$. It is obvious that the length over which the probe is coupled to the HES needs to be smaller than the equilibration scale dictated by $N_0$ so that the sensitivity of measured $V_R'$ to the spin-polarization of the HES survives, which is equivalent to saying that this will ensure that $V_R'$ contains a finite magnetoresistance contribution. Though it is clear that to measure spin-resolved voltages in a HES, we need a spin-polarized voltage probe, but to ensure that the probe indeed measures the spin-resolved voltages, we have to simultaneously optimize the tunneling strength and the physical extent of the coupling between the probe and the HES so that a complete equilibration of the spins on the HES induced by the probe does not obstruct the measurement of a spin-resolved $V_R'$. 

For the above study, we can conclude that the limit, in which an almost perfect readout of the spin-resolved voltage is possible, is when we are in the vicinity of $\theta=\pm\pi/2$ and the tunneling strength and the length of the entire junction (which is proportional to $N_P$) or alternatively, the physical extent of the coupling are chosen appropriately guided by Eq.~\ref{exp} so that we are always away form the regime of complete spin equilibration. Also, the case of  $\theta=\pm\pi/2$ with uniform $t'$ for the subprobes is too ideal to be realized in an experimental situation. This naturally compels us to consider the effects of fluctuations in $\theta$ and $t'$ across the subprobes in the next section where we analyze how such disorders would influence the readout of the SPVP. 

\subsection{Disordered case}

\begin{figure}
\centering
\includegraphics[width=1.0\columnwidth]{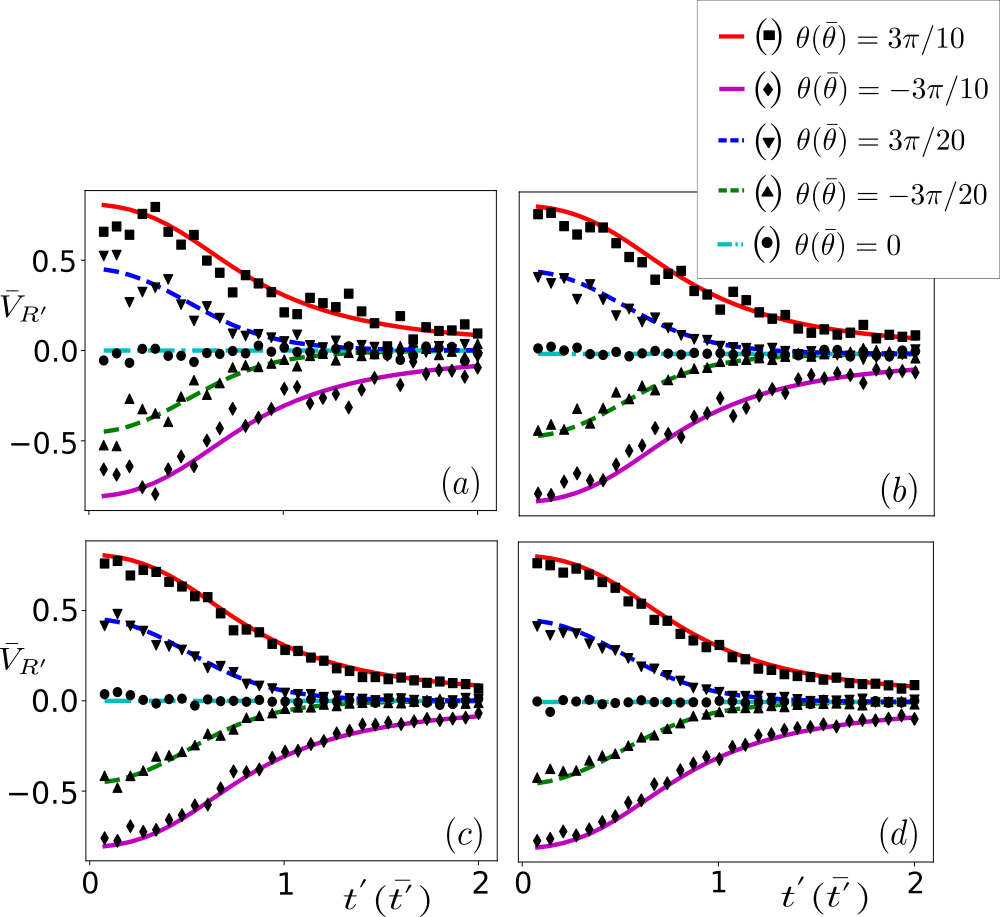}
\caption{Evolution of the disorder averaged variation of $\overbar{V}_{R'}$ as a function of the tunneling strength ($\overbar{t'}$) for different values of $\bar{\theta}$ with $N_P=10$. The solid lines in each figure represent the uniform case: the variation of ${V}_{R'}$ as a function of a uniform value of $t'$ for different values of  $\theta$. The scattered datapoints in (a) show the variation of $V_{R'}$ for a single random disorder configuration of $t'$ and $\theta$. Figure (b), (c), and (d) show the same variation after taking a disorder average over $5$, $10$, and $20$ random configurations of $t'$ and $\theta$ about their respective average values. We have taken $\sigma_t$ = $\sigma_{\theta} = 0.2$ and the offset along the $x$-axis of all the plots is taken to be $0.01$.}
\label{fig:figg3x}
\end{figure}

\begin{figure}
\centering
\includegraphics[width=1.0\columnwidth]{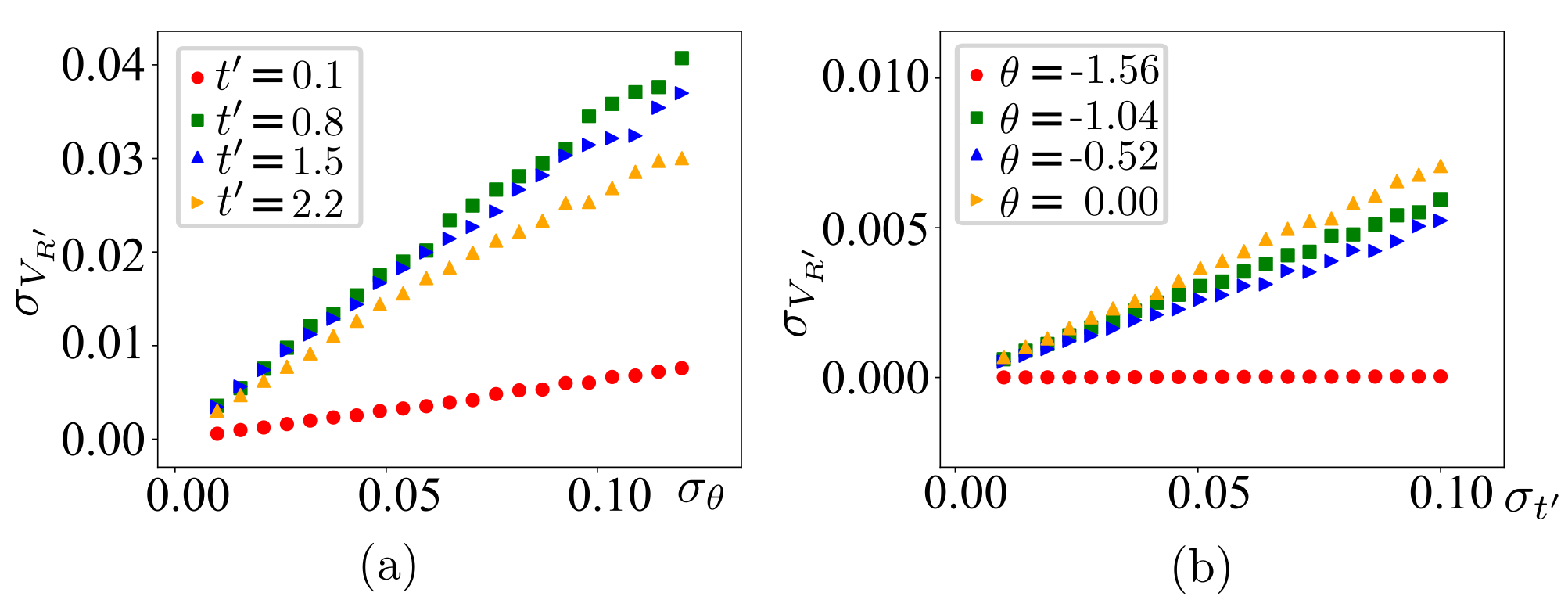}
\caption{
(a) Plot of $\sigma_{V_{R'}}$ vs $\sigma_{\theta}$ at different values of (uniform) $t'$ for a SPVP ($N_P=10$) with Gaussian disorder in $\theta$ across the subprobes with a mean value of $\theta$ to be $-1.27$. The fluctuations increase in the strong tunneling limit (averaging performed over $10^3$ realizations).(b) Plot of $\sigma_{V_{R'}}$ vs $\sigma_{t'}$ at different values of (uniform) $\theta$ for the SPVP ($N_P=10$) with Gaussian disorder in $t'$ across the subprobes with a mean value of $t'$ to be $1.0$. The fluctuations increase for the values of $\theta$ away from $\pm \pi/2$, but the magnitude is much smaller compared to the disordered $t'$ case shown in (a).}
\label{fig:figg3}
\end{figure}

In the disordered case, we first present a comparative study between different disorder averaging of $V_{R'}$ sensed by a SPVP of $N_P=10$, performed over ensembles of various number of disorder configurations of ${t'}$ and $\theta$. We denote the resultant voltage as $\overbar{V_{R'}}$ and plot it against the average value of the tunneling strength $\overbar{t'}$ for different values of the average polarization of the SPVP, $\overbar{\theta}$ in Fig.~\ref{fig:figg3x} for ensembles of size 1, 5, 10, and 20. As evident from Fig.~\ref{fig:figg3x}, we find that the averaging procedure results in a rapid convergence of the functional dependence of $\overbar{V_{R'}}$ to that for the uniform case as predicted by Eq.\ref{exp}; in fact, an ensemble of size as few as $20$ realizations is promising enough to obtain a close qualitative resemblance to the uniform case. 

Next we study the stability of the voltage measured by the SPVP as a function of fluctuations in the tunneling strength $t'$ and the polarization angle $\theta$ of the SPVP. In Fig.~\ref{fig:figg3} (a), we plot the fluctuations in $V_{R'}$ characterized by its standard deviation when the measurement is performed over many realizations of the subprobe disorders: $\sigma_{V_{R'}}=\sqrt{\overbar{V_{R'}^2}-\overbar{V_{R'}}^2}$, as a function of $\sigma_\theta$ for various values of the tunneling strength $t'$ which is assumed to be uniform across all subprobes (the disorder averaging is performed over $10^3$ realizations). Here $\sigma_\theta$  represents the standard deviation in the values of $\theta$ for various subprobes about the chosen average value of $\theta$. We note that the fluctuations in $V_{R'}$ increase monotonically with the increase in $\sigma_\theta$ as expected but the rate of increase is higher for larger values of the tunneling strength ($t'$), hence, indicating that small $t'$ limit is desirable for implementing such a voltage probe in presence of fluctuations in $\theta$.  Fig.~\ref{fig:figg3} (b) shows the variation of $\sigma_{V_{R'}}$ as a function of $\sigma_{t'}$ where $\sigma_{t'}$ represents the standard deviation in the tunneling strength $t'$ across various subprobes about a chosen average value of $t'$. If the value of $\theta$ corresponds to either the parallel or anti-parallel configuration ({\it i.e.}, $\pm\pi/2$), we expect $\sigma_{V_{R'}}$ to be strongly suppressed. We emphasize that in the plot corresponding to $\theta=-1.56$ (a value close to $\theta = -\pi/2$), increasing $\sigma_{t'}$ has a little influence on $\sigma_{V_{R'}}$ and it stays close to zero. But as we move away from these polarizations, $\sigma_{V_{R'}}$ displays a behavior which is similar to that observed against $\sigma_\theta$ discussed above. Also note that the rate of increase of $\sigma_{V_{R'}}$ as a function of $\sigma_{t'}$ is the steepest at $\theta=0$ as expected owing to the fact that at $\theta=0$, the magnetoresistance response of each subprobe will be completely suppressed. 

Hence, for the above study, we conclude that a strong suppression in $\sigma_{V_{R'}}$ will take place if we take the average value of $t'$ to be small and the average value of $\theta$ to be close to $\theta = \pm \pi/2$. In what follows in the next section, we carry forward the discussion and explore the possibility of using a SPVP for reading off the local spin-resolved voltage drops on a HES using the model  discussed above. In particular, we address a situation which is analogous to the six-probe Hall bar setup (two current probes and four voltage probes as in Fig.\ref{fig:figg4}~\cite{Ferry2015}) involving a quantum point contact (QPC) which represents an archetypal setup in the context of quantum Hall experiments. 

\section{Measuring Hall-type response in a six-probe setup for HES}
\label{secthree}

\begin{figure}
\centering
\includegraphics[width=1.0\columnwidth]{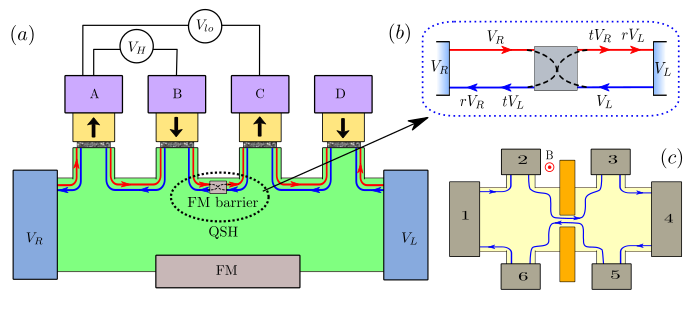}
\caption{(a) Schematic of the six-probe (four voltage probes and two current probes) setup to measure the Hall conductance between the counterpropagating states of the QSHS and the longitudinal conductance across an FM barrier (shaded grey, not to confuse with the FM placed on the bottom edge). The Hall voltage and the longitudinal voltage are denoted as $V_H$ and $V_{lo}$ respectively. The polarization angle for the SPVP are set as $\theta_A=\theta_C=-\pi/2=-\theta_B=-\theta_D$ such that SPVP A and C sense the voltage on the $\uparrow$-channel (right movers) and SPVP B and D sense the voltage on the $\downarrow$-channel (left movers) only. A ferromagnet (FM) is employed to open a gap in the spectrum of the bottom edge hence, electrically disconnecting the left and the right leads via this edge. (b) The boxed figure on the right top shows the zoomed-in FM barrier with a transmission probability of $t=91.39\%$ and the voltage drops across this barrier depending on $t$ and the reflection probability $r=1-t$  that characterize the scattering through the barrier. (c) The analogous schematic of a six-probe Hall bar setup involving a QPC in the middle as discussed in Ref.~\onlinecite{Ferry2015} where lead 1 and 2 represent the current probes and lead 3, 4, 5, and 6 represent the voltage probes.
}
\label{fig:figg4}
\end{figure}

Consider the FM barrier, localized at $x=x_0$ on the HES, is described by the Hamiltonian
\begin{align}
{\cal H}_{\rm FM} = \int_{-\infty}^{\infty} dx~ \delta(x-x_0) \big[{\cal B} \psi_R^\dagger \psi_L + {\rm h.c.} \big],
 \label{hcoup}
\end{align}
where $|{\cal B}|$ represents the coupling strength (proportional to the in-plane magnetic field produced by the FM barrier) and the  Hamiltonian for the HES is already given in Eq.~\ref{Hedge}. In general, ${\cal B}$ can be a complex number, however, the transfer matrix that connects the currents across the FM barrier is independent of the phase of ${\cal B}$, and thus, $\pm|{\cal B}|$ is the only relevant input that enters the calculations for voltages measured by the probe. Hence, without loss of generality, we consider ${\cal B}$ to be real. The currents across the FM barrier are related as
\begin{align}
 I_R(x_0^+) &= T^{\rm (FM)}_{RR}I_R(x_0^-) + T^{\rm (FM)}_{RL}I_L(x_0^-) \nonumber \\
 I_L(x_0^+) &= T^{\rm (FM)}_{LR}I_R(x_0^-) + T^{\rm (FM)}_{LL}I_L(x_0^-),
\end{align}
where the lower index of $R, L$ represents the right and the left movers; $x_0^{\pm}= x_0 \pm \epsilon$, $\epsilon$ being a vanishingly small positive number. The corresponding transfer matrix elements for the currents are
\begin{align}
 T^{\rm (FM)}_{RR} &= |s^{\rm (FM)}_{RR}|^2-|s^{\rm (FM)}_{RL}|^2|s^{\rm (FM)}_{LR}|^2/|s^{\rm (FM)}_{LL}|^2 \nonumber \\
 T^{\rm (FM)}_{RL} &= s^{\rm (FM)}_{RL}|^2/|s^{\rm (FM)}_{LL}|^2 \nonumber \\
 T^{\rm (FM)}_{LR} &= -s^{\rm (FM)}_{LR}|^2/|s^{\rm (FM)}_{LL}|^2 \nonumber \\
  T^{\rm (FM)}_{LL} &= 1/|s^{\rm (FM)}_{LL}|^2, 
\label{FB}
\end{align}
$s^{\rm (FM)}_{ab}$ being the elements of the scattering matrix for the FM barrier (full expressions given in Appendix~\ref{appB}). 

\begin{figure}
\centering
\includegraphics[width=1.0\columnwidth]{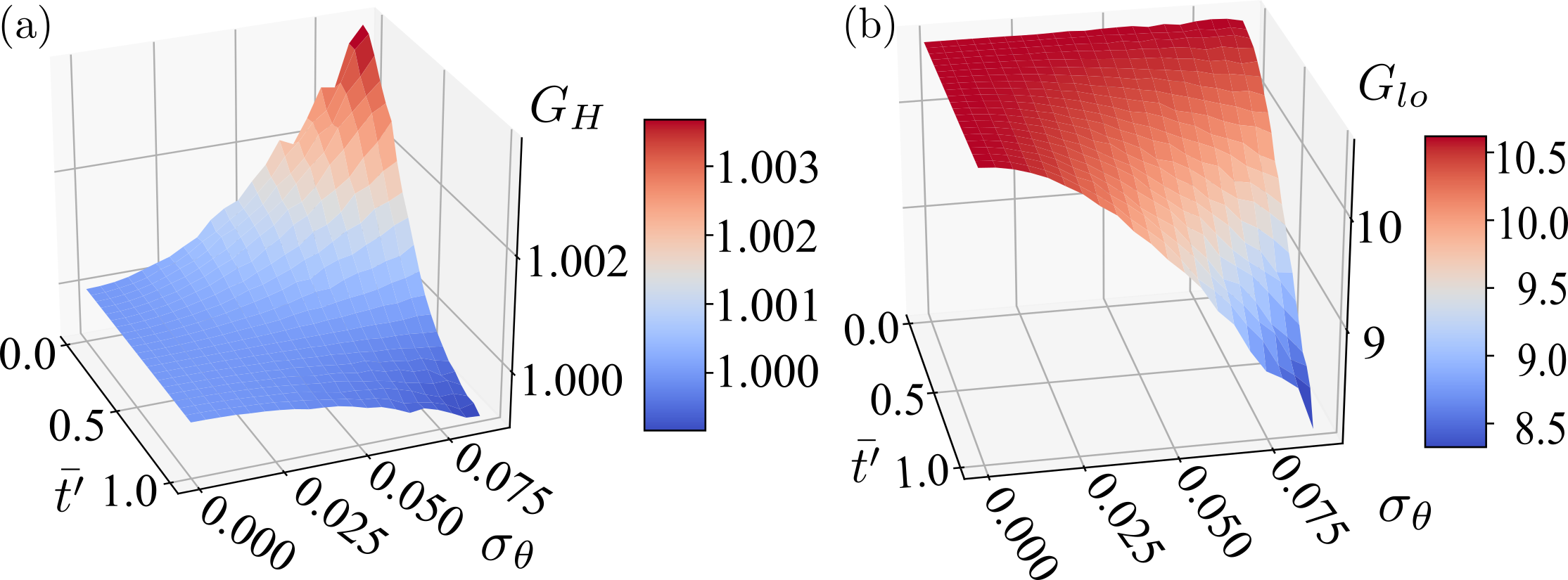}
\caption{Plot of the Hall conductance $G_H$ in (a) and the longitudinal conductance $G_{lo}$ in (b) as a function of $\bar{t'}$ and $\sigma_\theta$ (defined in the main text). The variation of $G_H$ being small signifies its robust nature against disorder as illuminated in the main text.}
\label{fig:figg6}
\end{figure}

Let us first consider the situation in which all the subprobes in each of the SPVPs, denoted by A, B, C, and D in Fig.~\ref{fig:figg4} (a), are coupled to the HES with uniform tunneling strength ($t'$) with equal $N_P$ for each of them. The polarization angle of the SPVPs are further set according to $\theta_A=\theta_C=-\theta_B=-\theta_D=-\pi/2$ such that SPVP A and C measures the voltage for the $\uparrow$-channel only  (note we have considered a shift in $\theta$ by $\pi/2$ at the starting of our formulation) and SPVP B and D do so for the $\downarrow$-channel only (see Eq.~\ref{probvolt2}). In Fig.~\ref{fig:figg4} (c) we have shown a Hall bar geometry where the Hall voltage can be defined as the voltage difference measured between the voltage probes $2, 6$ or $3, 5$. In the setup for the HES in Fig.~\ref{fig:figg4} (a), the analog of Hall voltage $V_H$ is represented by the voltage difference between the $\uparrow$-channel and the $\downarrow$-channel on the same side of the FM barrier, $V_H=V_R-(rV_R+tV_L)=t(V_R-V_L)$ where the FM barrier plays the role of the QPC in the Hall bar geometry given in  Fig.~\ref{fig:figg4} (c) and further demonstrated in Fig.~\ref{fig:figg4} (b). The net current in the HES is $I=tV_R+rV_L-V_L=t(V_R-V_L)$ and hence, the Hall resistance can be defined as  $R_H=V_H/I=1$ (in the units of $e^2/h$) which is quantized~\cite{MacDonald1973Edge}. The longitudinal voltage drop $V_{lo}$ is the voltage difference along a given spin-polarized edge ($\uparrow$ or $\downarrow$) across the barrier which is given by $V_{lo}=V_R-(tV_R+rV_L)=r(V_R-V_L)$ and so, the longitudinal resistance is given by $R_{lo}=V_{lo}/I=r/t$ which is nothing but the four-probe resistance~\cite{Yoseph2002} for the FM barrier. The sum of these two resistances satisfies $R_{lo}+R_H =1/t$ which is expected from the Landauer formula~\cite{LandauerIBM1957}. 

Now we will perform a self consistent numerical analysis to implement the setup described above in terms of our model for the SPVP and check if it reproduces the expected behaviors of $R_H$ and $R_{lo}$. The numerical calculations are performed considering the coupled equations for the subprobes given in Eq.~\ref{curr_mat3} for all the four SPVPs. These equations (each for SPVP A, B, C, and D) are further coupled to each other via transfer matrix for the current which, in the case between SPVP A and SPVP B, and between SPVP C and SPVP D, is an identity matrix while between SPVP B and SPVP C, is the transfer matrix for the FM barrier given in Eq.~\ref{FB}. Finally, the voltage probe condition is imposed simultaneously on SPVP A, B, C, and D and the full set of equations are solved numerically. As fluctuations in both $t'$ and $\theta$ for each of the SPVPs are expected to exist in a realistic situation, we perform our numerical analysis in their presence and check for the stability in the obtained value of $V_{R'}$ for each probe.

As discussed in the previous section, we model such fluctuations with a Gaussian distribution of $\theta$ among the subprobes in each of the SPVPs with the mean $\overbar{\theta_A}=\overbar{\theta_C}=-\overbar{\theta_B}=-\overbar{\theta_D}=-\pi/2$ and standard deviation $\sigma_\theta$ ranging from $10^{-4}$ to $\pi/36$ ($\sim 5.56\%$ of $\pi/2$). We further include such Gaussian disorder in the tunneling strength $t'$ as well with a mean $\overbar{t'}\in [0.01,1]$ and standard deviation $\sigma_{t'}=0.01$ in each of the SPVPs. For the coupling strength of the barrier, ${\cal B}=0.3$, which corresponds to a transmission probability of $t=91.39 \%$ (using Eq.~\ref{bs3} in the Appendix), we study the Hall conductance $G_{H}$ (this involves voltage difference between SPVP A and B) and the longitudinal conductance $G_{lo}$ (involving voltage difference between SPVP A and C) as a function of $\overbar{t'}$ and $\sigma_{\theta}$ following a disorder averaging over $10^3$ realizations. The results are shown in Fig.~\ref{fig:figg6}. 

For the FM barrier with transmission probability of $91.39 \%$, the theoretical values for $G_{H}=1$ and $G_{lo}=t/r\approx 10.6$. We note from Fig.~\ref{fig:figg6} that the our numerical analysis reproduces these predicted theoretical values very well in the limit of $\sigma_{\theta} \rightarrow 0$ independent of the value of $\overbar{t'}$ where  $\sigma_{t'}$ is taken to be $0.01$. The plot illuminates the robust nature of the disorder-averaged value of $G_{H}$ which shows significantly small variations in its value when plotted as a function of $\sigma_{\theta}$. An increase in $\overbar{t'}$ tends to stabilize the value of $G_{H}$ and forces it stay close to the value corresponding to $\sigma_\theta=0$. But we must keep in mind that the standard deviation in $G_{H}$ should also increase as we increase $\overbar{t'}$ which is known from the results presented in Fig.~\ref{fig:figg3}, {\it i.e.}, $\sigma_{V_{R'}}$ increases with increasing $\sigma_{\theta}$ at a faster rate for larger values of $\overbar{t'}$ and hence, $G_{H}$ can develop large fluctuations due to its dependence on $V_{R'}$. Hence, though the average $G_{H}$ stays stable with $\overbar{t'}$, the fluctuations in its values tend to increase in this limit implying that it will be safe to operate in the small $\overbar{t'}$ limit for reproducing the theoretically expected results. 

\begin{figure}
\centering
\includegraphics[width=1.0\columnwidth]{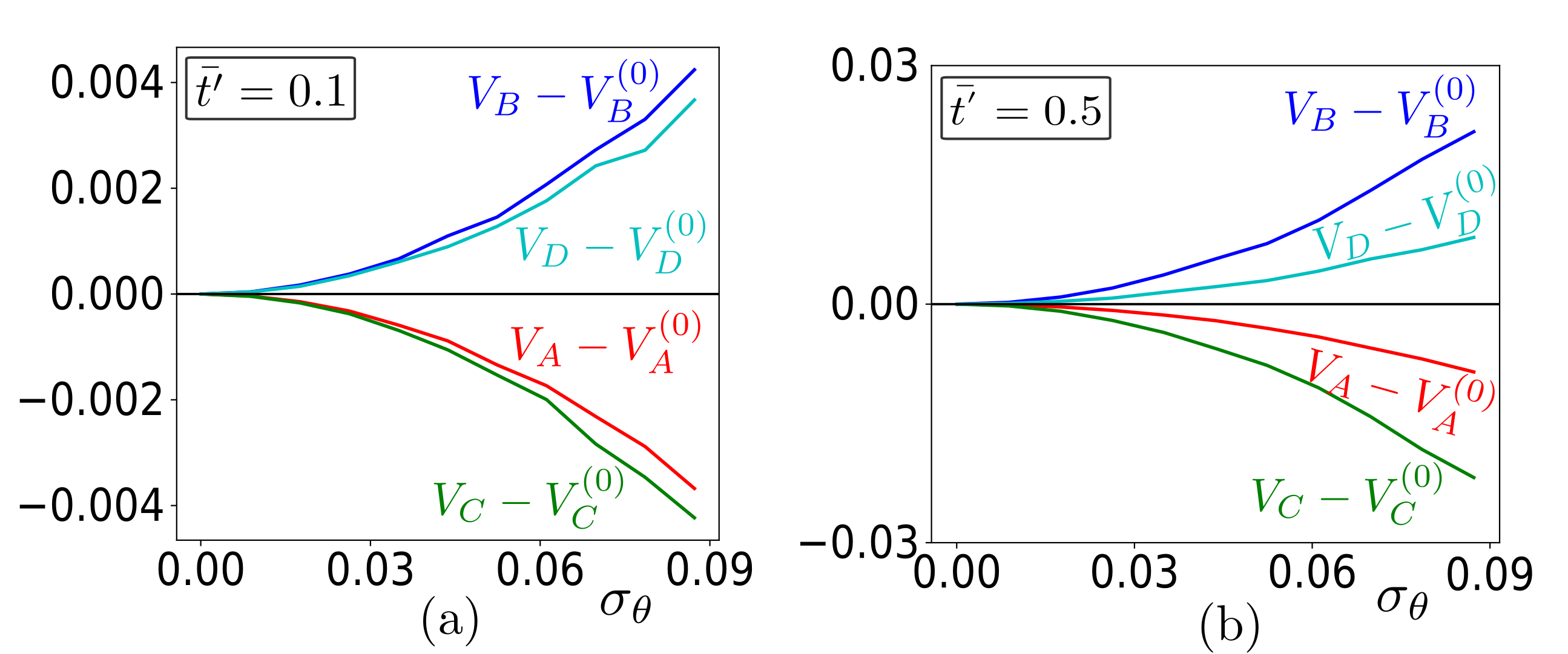}
\caption{Plot of individual probe voltages with respect to their values for the uniform case, {\it i.e.}, $V_m-V_m^{(0)}$ ($m=A,B,C,D$), as function of $\sigma_\theta$ for $\overbar{t'}=0.1$ in figure (a) and $\overbar{t'}=0.5$ in figure (b). A disorder averaging has been performed over 200 realizations in each of the plots.}
\label{fig:figg7}
\end{figure}

For the longitudinal conductance $G_{lo}$, we observe approximately a constant value as we increase disorder in the polarization angle ($\sigma_\theta$) at small values of $\bar{t'}$, however, the variation increases with increasing $\bar{t'}$ as seen in Fig.~\ref{fig:figg6} (b). This can be understood better by inspecting the output voltages in the individual SPVPs (measured with respect to their values for the uniform case, {\it i.e.}, the disorder free case) as plotted in Fig.~\ref{fig:figg7} at different values of $\bar{t'}$. At very small values of $\bar{t'}$ [{\it e.g.} $\bar{t'}=0.1$ as in Fig.~\ref{fig:figg7} (a)], the profile of $V_m-V_m^{(0)}$ ($V_m^{(0)}$ denoting the value for the disorder free case) for $m=A$ and $m=C$ (and likewise, $m=B$ and $m=D$) almost match resulting in small variation in $G_{lo}$ as observed in Fig.~\ref{fig:figg6} (b) while it can have an adverse effect on $G_{H}$ though small [Fig.~\ref{fig:figg6} (a)]. But with increasing $\bar{t'}$ [{\it e.g.} $\bar{t'}=0.5$ or higher as in Fig.~\ref{fig:figg7} (b)], the profile of $V_A-V_A^{(0)}$ and $V_C-V_C^{(0)}$ (and likewise, $V_B-V_B^{(0)}$ and $V_D-V_D^{(0)}$) tend to move away from each other resulting in $G_{lo}$ to fall off steeply from its uniform value with increasing $\sigma_\theta$. 

To conclude, we note that the small $\bar{t'}$ limit is very important for implementing multi-terminal spin-resolved voltage measurements in presence of disorder ({\it i.e.}, finite $\sigma_\theta$ and $\sigma_{t'}$) in the SPVPs with $N_P\gg1$. 
Of course in the ideal limit of $N_P=1$, the value of ${t'}$ is of no consequence as long as 
${\theta_A}={\theta_C}=-{\theta}_B=-{\theta_D}=-\pi/2$ but that will correspond to an unrealistic situation as in any ohmic contact tunnel-coupled to HES, which could be used as a voltage probe, is expected to host multiple modes owing to its finite size. 

\section{Spin polarized chiral edge state as a SPVP}
\label{secfour}
\begin{figure}
\centering
\includegraphics[width=1.0\columnwidth]{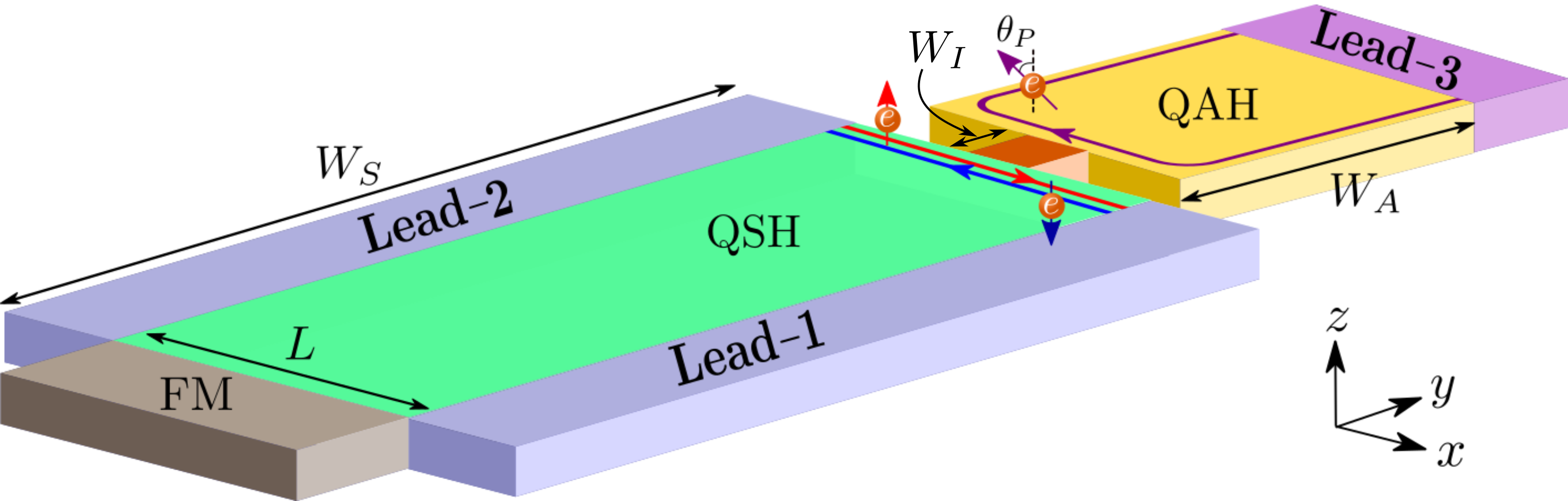}
\caption{Model of a voltage probe consisting of a single chiral edge acting as a SPVP (constructed from a QAH system) tunnel-coupled to the HES of a QSHS as used in the simulation ($L$, $W_S$, and $W_A$ specified in the main text). Lead-1 and Lead-2 represent the leads which act as the electron reservoirs used to apply a finite voltage bias across the HES while Lead-3 is used as a voltage probe with probe voltage given by $V_{R'}$. The FM is employed to open a finite gap in the spectrum of the electronic states lying on the lower HES. A square insulating region of dimension $W_I$ (specified in the main text) is used to facilitate a controlled tunneling between the QAH edge and the HES. The polarization of the HES is along $S_z$ while that of the QAH edge is tilted from $S_z$ at an angle $\theta_P$ which can be tuned continuously.}
\label{fig:figg8}
\end{figure}

In the study presented above, we have considered a theoretical model where the SPVP comprises a number of chiral one-dimensional spin-polarized modes which are mutually incoherent with each other and are tunnel-coupled to the HES via local tunneling. In this section, we consider the possibility of using a simpler model for the SPVP where a single spin-polarized chiral edge is employed for the purpose and the tunnel-coupling between the spin-polarized chiral edge and the HES is taken to be spread out over an extended region rather than being local which makes it complementary to our previous model. We simulate a lattice model involving a SPVP coupled to the HES of a QSHS using the software package KWANT~\cite{groth2014kwant}. The full lattice model consists of three parts (schematically shown in Fig.~\ref{fig:figg8}): {\it (i)} a quantum spin Hall (QSH) region described by the BHZ model~\cite{bernevig2006quantumnature} in its topological phase hosting HES at the edge, {\it (ii)} a quantum anomalous Hall (QAH) region described by the BHZ model with an applied exchange field which drives it into the QAH phase that supports a single spin-polarized chiral edge mode whose polarization can be controlled via rotating the direction of the applied field and, {\it (iii)} an insulating barrier between the QAH edge and the HES which is described by the BHZ model in its trivial phase. The dimensions of the QSH region are taken to be $L\times W_S$ with $L=100a$, $W_S=147a$, and $a=3$ nm being the lattice constant. The dimensions of the QAH region is taken as $L\times W_A$ with $W_A=50a$, and that of the insulating region is taken to be of dimensions $3a\times 3a$. 

The bulk of the QSH region is given by the BHZ Hamiltonian
\begin{align}
H_{\rm BHZ} = -D k^2 + A k_x \sigma_z \tilde{\sigma}_x - A k_y \tilde{\sigma}_y + (M-B k^2) \tilde{\sigma}_z,
 \label{BHZham}
\end{align}
where $\sigma$ and $\tilde{\sigma}$ denote the Pauli matrices to describe the spins ($\uparrow$ or $\downarrow$ along $S_z$) and the orbitals ($s$ or $p$ type) respectively and, $A$, $B$, $D$, and $M$ are material-dependent parameters. For HgTe/CdTe quantum wells, $D = -512.0~{\rm nm}^2$-meV, $A = 364.5$ nm-meV, $B = -686.0~{\rm nm}^2$-meV, and $M = -10.0$ meV, however, in our simulation, we set $D =0$ to place the Dirac point at zero energy. A discrete form of the Hamiltonian in Eq.~\ref{BHZham} can be identified on a square lattice with a basis of two sites (representing the two orbitals) using $k^2 = 2 a^{-2}[2-\cos(k_x a)-\cos(k_y a)]$, $k_x = a^{-1} \sin(k_x a)$, and $k_y = a^{-1} \sin(k_y a)$ ($a$ being the lattice constant specified before), which reads~\cite{chen2016pi, adak2020spin}
\begin{equation}
H_{\rm tb} = \sum_i (c_i^{\dagger} H_{i,i+a_x} c_{i+a_x} + c_i^{\dagger} H_{i,i+a_y} c_{i+a_y} + {\rm h.c.}) + c_i^{\dagger} H_{ii} c_i,
\end{equation}
where $c_i^{\dagger} \equiv (c^{\dagger}_{i,s,\uparrow},c^{\dagger}_{i,p,\uparrow},c^{\dagger}_{i,s,\downarrow},c^{\dagger}_{i,p,\downarrow})$ denotes the set of creation operators for the electrons in $s$ and $p$ orbital with $\uparrow$ and $\downarrow$ spins at site $i$ with coordinates $i=(i_x,i_y)$; $a_x = a(1,0)$ and $a_y = a(0,1)$ are the lattice vectors. Each of the terms, $H_{ii}$ and $H_{i,i+a_x(a_y)}$, is a $4 \times 4$ block matrices defined by
\begin{align}
H_{ii} &= -\frac{4 D}{a^2} - \frac{4 B}{a^2} \tilde{\sigma}_z + M \tilde{\sigma}_z \nonumber \\
H_{i,i+a_x} &= \frac{D + B \tilde{\sigma}_z}{a^2} + \frac{A \sigma_z \tilde{\sigma}_x}{2 i a} \nonumber \\
H_{i,i+a_y} &= \frac{D + B \tilde{\sigma}_z}{a^2} + \frac{i A \tilde{\sigma}_y}{2 a}.
\label{bhzlattice}
\end{align}
The lattice constant is set to $a =3$ nm to obtain a realistic band structure. As mentioned earlier, the spin-polarized chiral edge, that models the SPVP in this case, is obtained from the QSHS by inducing a topological phase transition via the application of an exchange field of strength $g_0$ in the BHZ Hamiltonian ( $|g_0|> |M|$~\cite{liu2008quantum, li2013stabilization} ), where the direction of spin-polarization of the chiral edge can be tuned by tuning the direction of the applied exchange field and replacing the coefficient of $A \, k_x$ term with an appropriate term as done below. Hence the model Hamiltonian reads as
\begin{align}
\tilde{H}_{\rm BHZ} \nonumber = -D k^2 + A k_x ({\hat{\bf a}}.{\bf \sigma}) \tilde{\sigma}_x - A k_y \tilde{\sigma}_y &+ (M-B k^2) \tilde{\sigma}_z \\
&+ g_0 ({\hat{\bf a}}.{\bf \sigma}) \tilde{\sigma}_z,
 \label{BHZano}
\end{align}
where we have replaced the term $A k_x \sigma_z \tilde{\sigma}_x$ in Eq.~\ref{BHZham} with a new term $A k_x ({\hat{\bf a}}.{\bf \sigma})\tilde{\sigma}_x$ such that $S_{\bf a}$ (the component of spin along $\hat{\bf a}$) is conserved  and the  unit vector ${\hat{\bf a}}$ deciding the new direction of spin-polarization on the edge. 
The corresponding lattice model is given by 
\begin{align}
\tilde{H}_{ii} &= -\frac{4 D}{a^2} - \frac{4 B}{a^2} \tilde{\sigma}_z + M \tilde{\sigma}_z + g_0 ({\hat{\bf a}}.{\bf \sigma}) \tilde{\sigma}_z \nonumber \\
\tilde{H}_{i,i+a_x} &= \frac{D + B \tilde{\sigma}_z}{a^2} + \frac{A ({\hat{\bf a}}.{\bf \sigma}) \tilde{\sigma}_x}{2 i a} \nonumber \\
\tilde{H}_{i,i+a_y} &= \frac{D + B \tilde{\sigma}_z}{a^2} + \frac{i A \tilde{\sigma}_y}{2 a}.
\label{bhzlatticeano}
\end{align}
In our case, we consider $g_0=-15$ meV in the QAH region which, for $\theta_P=0$, yields a chiral $\uparrow$ edge in the QAH region and for $\theta_P=\pi$, a chiral $\downarrow$ edge. A lead (denoted as Lead-3 in Fig.~\ref{fig:figg8}) consisting of the same Hamiltonian as in Eq.~\ref{bhzlatticeano} is attached to the bottom edge (the edge away from the QSH region) of the QAH region through which the spin-resolved probe voltage $V_{R'}$ could be measured as a function of the polarization angle $\theta_P\equiv\cos^{-1}({\hat{\bf a}}.{\bf \sigma})\in[0,\pi]$ as described below.
The insulating region is a trivial insulator, characterized by the same Hamiltonian as in Eq.~\ref{bhzlatticeano}, however, $g_0=0$ and the sign of the mass term $M$ flipped (taken $M = 10.0$ meV) in this region as required for driving a QSHS into a trivial insulating state. 

\begin{figure}
\centering
\includegraphics[width=1.0\columnwidth]{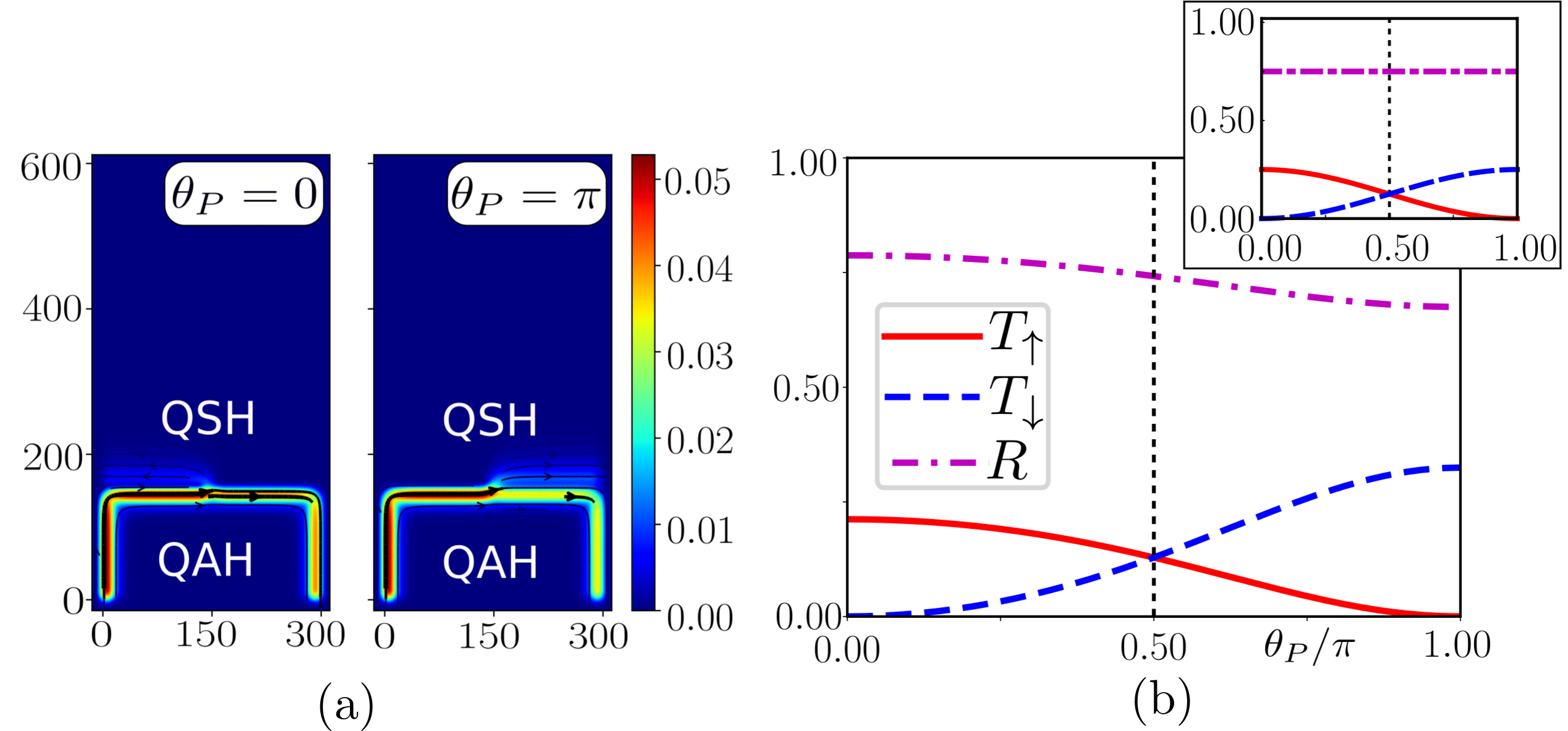}
\caption{(a) Plot of the current density on the entire lattice used in the simulation at two values of the probe polarization angle $\theta_P$. The tick labels in the figure are given in nanometers. The bright regions indicate the presence of the (chiral) current with its flow guided by the black arrows. (b) Plots of $T_\uparrow$, $T_\downarrow$, and $R$ (defined in the main text) as function of $\theta_P$ as obtained from the numerics for an incident energy of $E=10^{-3}$ meV. The inset shows the ones obtained from an analytical calculation (details mentioned in the main text).}
\label{fig:figg9}
\end{figure}

With this setup, we first study the current density which is proportional to the transmittance from the chiral (QAH) edge to the helical edge of the QSH region, denoted $T_{\uparrow}$ and $T_\downarrow$ for the $\uparrow$-channel and $\downarrow$-channel respectively (the reflectance along the chiral channel is $R=1-T_\uparrow-T_\downarrow$). We plot the current density in Fig.~\ref{fig:figg9} (a) for two values of $\theta_P$ when electrons are incident at energy $E=10^{-3}$ meV which clearly indicates the spin-polarization direction of QAH edge with respect to the two spin-momentum locked modes of the HES. As mentioned above, at a given value of $\theta_P$, the spin-polarization of the chiral edge has components along that of both the channels of the helical edge resulting in transmission into the $\uparrow$-channel with magnitude $T_{\uparrow}$ and into the $\downarrow$ channel with magnitude $T_{\downarrow}$. An interesting feature to note in this plot is that when $\theta_P$ is changed to $\pi$ from $0$, the positions of the two blocks characterized by $\sigma_z=\pm1$ (the eigenvalues of $\sigma_z$ being good quantum numbers as $[\sigma_z,\tilde{H}_{\rm BHZ}]=0$ at $\theta_P=0,\pi$) in the modified BHZ Hamiltonian $\tilde{H}_{\rm BHZ}$ (Eq.~\ref{BHZano}) are switched, and as a result, when $g_0$ is applied with an appropriate magnitude to counter the mass term $M$, the spin on the (chiral) edge that survives gets flipped, however, the chirality is retained. This is evident from the direction of the current density (black arrows) in both the plots at $\theta_P=0$ and $\theta_P=\pi$. The spin of the chiral (QAH) edge is flipped for $\theta_P:0\rightarrow \pi$ and consequently the transmittance changes from $T_\uparrow(\theta_P=0)\neq 0, T_\downarrow(\theta_P=0)= 0$ to $T_\uparrow(\theta_P=\pi)= 0, T_\downarrow(\theta_P=\pi)\neq 0$ as shown in Fig.~\ref{fig:figg9} (b). The zeros in the transmission probabilities demonstrate a perfect suppression in electron tunneling between the spin orthogonal modes. The plot in Fig.~\ref{fig:figg9} (b) shows the full behavior of $T_\uparrow$, $T_\downarrow$, and $R$ as function of $\theta_P$. The inset displays the behaviors of $T_{\uparrow}$ , $T_{\downarrow}$, and $R$ for the model depicted in Fig.~\ref{fig:figgxxx} for a single subprobe with a local tunnel-coupling between the HES and the subprobe obtained analytically from the scattering matrix at a tunneling strength of $t'=0.5$ (details given in Appendix~\ref{appA}). Note that our two-dimensional transport simulation shows a close  qualitative resemblance to the ideal model for the subprobe discussed in the previous section. 

The discrepancy between the results obtained for lattice simulation and the one obtained from a simple minded analytical model given in Eq.~\ref{probvolt2} can be minimized by adjusting the size of the insulating region and the plots shown above in Fig.~\ref{fig:figg9} are the closest we could obtain that exhibit a slight variation of $R$ against $\theta_P$ (albeit about a mean value that matches the analytics). Note that the simple-minded one-dimensional model for the subprobe assumes a tunnel-coupling which respects spin-rotation symmetry and this fact leads to the symmetric variation of $T_{\uparrow}$ and $T_{\downarrow}$ about $\theta_P=\pi/2$. But, on the other hand, the presence of finite spin-nonconserving tunneling events at the junction between the SPVP and the HES owing to the presence of an applied exchange field at the microscopic level in the lattice Hamiltonian constituting the junction always leads to the observed asymmetry in the main figure of Fig.~\ref{fig:figg9} (b).

\begin{figure}
\centering
\includegraphics[width=1.0\columnwidth]{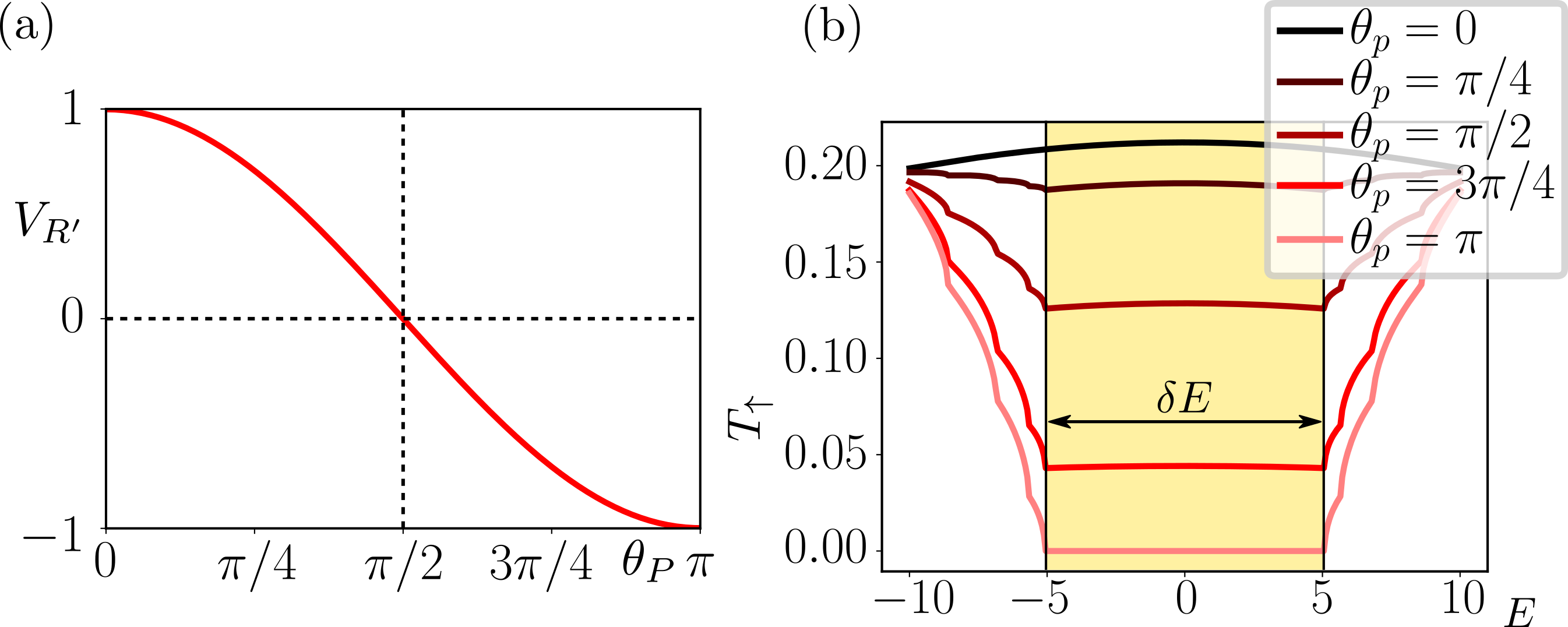}
\caption{(a) Plot of the probe voltage $V_{R'}$ vs. $\theta_P$ as calculated from Eq.~\ref{probvolt4} showing its variation from $V_{R'}=V_R=1$ at $\theta_P=0$ to $V_{R'}=V_L=-1$ at $\theta_P=\pi$ as expected. (b) A plot of $T_\uparrow$ vs. the incident energy $E$ at different values of $\theta_P$ reveals an energy window $E\in \delta E=[-5,5]$ meV within which the spin-resolved voltage measurement should be performed.}
\label{fig:figg10}
\end{figure}

Now let us consider a situation where $V_R>V_L$. Then the incoming current flowing into the probe is $I_{\uparrow}=T_{\uparrow}(V_R-V_{R'})$ while the outgoing one is $I_{\downarrow}=T_{\downarrow}(V_{R'}-V_L)$. Assuming that the transmission probabilities $T_{\uparrow}$ and $T_{\downarrow}$ are independent of energy in the bias window $V_R-V_L$, the voltage probe condition, implying a zero net current in the probe, leads to 
\begin{align}
 V_{R'} = \frac{T_\uparrow V_R + T_\downarrow V_L}{T_\uparrow + T_\downarrow}=\frac{V_R+ V_L}{2} +\frac{1}{2}\frac{T_\uparrow - T_\downarrow }{T_\uparrow + T_\downarrow} (V_R-V_L)  ,
 \label{probvolt4}
\end{align}
similar to Eq.~\ref{probvolt1}, which we plot against $\theta_P$ in Fig.~\ref{fig:figg10} (a). Note that the expression for $V_{R'}$ has two contributions: the first one is the average voltage of the left and the right moving edge which would have been the full contribution in absence of electron spin, while the second term, being proportional to $(V_R-V_L)$, is indeed the magnetoresistance contribution. In the plot presented in Fig.~\ref{fig:figg10} (a), $V_R=-V_L$ and hence, it features only the magnetoresistance response. 

Finally, it is important to identify an energy window in which the transmission probabilities have a very weak energy dependence for the SPVP to work and hence, we try to check for the presence of such an energy window in our lattice simulations. This is expected to work as we are working with edge states whose spectrum can be approximated to be linear to a large extent. We provide an estimate of an energy window $\delta E$ for our setup such that $V_{R/L}\in [-\delta E/2,\delta E/2]$ leads to treating $T_\uparrow$ and $T_\downarrow$ as constants under the variation of $E$. This window is shown in Fig.~\ref{fig:figg10} (b) by the yellow shaded region in the plot of $T_\uparrow$ against the incident energy $E$ at various values of $\theta_P$. This plot clearly suggests that, within the band gap $\delta E$, $T_\uparrow$ and $T_\downarrow$ can indeed be approximated as constants but beyond this gap, the transmittances largely deviate from their constant values due to the contributions of the bulk bands gradually coming into play.
\begin{figure}
\centering
\includegraphics[width=1.0\columnwidth]{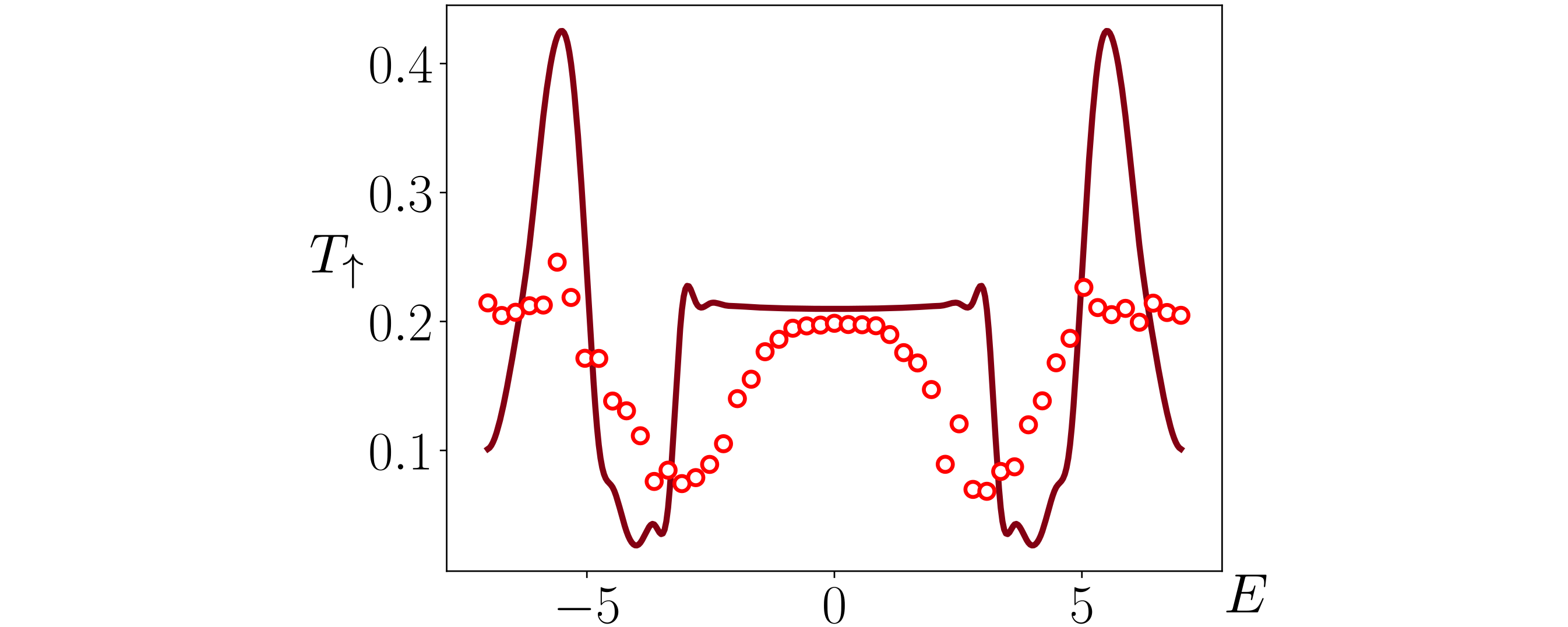}
\caption{Plot of $T_\uparrow$ (empty circles) vs. the incident energy $E$ in presence of disorder in the KWANT setup involving the InAs/GaSb system (details in the text). The disorder average is performed over 50 configurations and the uniform case (absence of disorder) is shown by the solid line.}
\label{fig:figg13}
\end{figure}

To account for the effect due to disorder in the SPVP, a modified BHZ model is simulated which describes the QSH state in an inversion symmetry broken system such as InAs/GaSb quantum well. The Hamiltonian, in this case,  accommodates spin-orbit interaction which breaks spin rotation symmetry about $S_z$~\cite{mi2013proposal} 
\begin{align}
H_{\rm SO} &=  c_0 \sigma_y \tilde{\sigma}_y + k_x \Big[\frac{c_+}{2} \sigma_x  + \frac{c_-}{2} \sigma_x \tilde{\sigma}_z - \frac{c_3}{2} \sigma_y (\mathbbm{1} + \tilde{\sigma}_z)\Big] \nonumber \\
&- k_y \Big[\frac{c_-}{2} \sigma_y + \frac{c_+}{2} \sigma_y \tilde{\sigma}_z - \frac{c_3}{2} \sigma_x (\mathbbm{1} + \tilde{\sigma}_z)\Big],
\end{align}
where, $c_{\pm} = c_1 \pm c_2$, in addition to $H_{\rm BHZ}$ in Eq.~\ref{BHZham} but including the parameter $g_0$ and a chemical potential term ($U$) as
\begin{align}
 H_{\rm BHZ} &= - D k^2 + A k_x \sigma_z \tilde{\sigma}_x - A k_y \tilde{\sigma}_y + (M-B k^2) \tilde{\sigma}_z \nonumber \\
 &~~~~~~~~~~~~~~~~~~~~~~~~~~~~~~~~~~~~~~~~~~~~~~~~~~ + g_0 \sigma_z \tilde{\sigma}_z + U.
\label{BHZg0}
\end{align}
The chemical potential $U$ takes random values on the lattice chosen from a uniform distribution with an energy cutoff: $-E_{\rm c}\leq U \leq E_{\rm c}$ where $E_{\rm c}$ can at most be equal to the bulk gap. A weak magnetic field ${\bf B}=(0,0,B_0)$ is allowed to facilitate breaking of time reversal symmetry, however, the corresponding Zeeman energy is neglected. The magnetic field enters the Hamiltonian only via the Peierls phase ${\rm Exp}[i(e/\hbar)\int {\bf A}\cdot{\rm d}{\boldsymbol\ell}]$ of the hopping terms specified by the gauge potential ${\bf A}=(0,B_0 x,0)$. For the InAs/GaSb system used in the simulation, the following parametric values are used: $A$ = 37.0 meV-nm, $D = 0.0$ meV-nm$^2$, $B = -660.0$ meV-nm$^2$, $M = -7.8$ meV, $g_0 = -12$ meV, $B_0 = 0.05$ T, $c_0 = 0.2$ meV, $c_1 = 0.066$ meV-nm, $c_2 = 0.06$ meV-nm, $c_3 = -7.0$ meV-nm, and $E_{\rm c}=7$ meV. We note that to demonstrate prominent effects of disorder, we have considered the insulating region to be described by the HgTe system however its dimensions are $3a\times9a$ where $a=10$ nm. The other relevant dimensions of the lattice as shown in Fig.~\ref{fig:figg8} are $L = 100a, W_S = 141a$, and $W_A = 50a$.

The results are shown in Fig.~\ref{fig:figg13} which displays a plot of the transmission probability $T_\uparrow$ (for the probe polarization $\theta_P=0$) vs the incident energy $E$ in presence of a disordered configuration of the chemical potential $U$, averaged over $50$ such disorder configurations. In the absence of disorder, a plateau at $T_\uparrow\approx 0.2$ is visible over an energy window $-4\leq E\leq 4$ meV, similar to what is observed also for the HgTe system [see Fig.~\ref{fig:figg10} (b)]. When disorder is taken into account, $T_\uparrow$ shows a stable behavior, however, the plateauing transpires over a smaller energy window. This makes us conclude that the SPVP constructed from the chiral edge of the QAH system can yield robust measurements of spin-resolved voltages on a HES in presence of disorder as well.    

\section{Conclusion}\label{secfive} 
Spin-momentum locked spectrum of surface states of two and three-dimensional topological insulators is a resource for a variety of spintronics applications. Though these states have no net polarization in equilibrium, once a finite bias is applied, they do develop finite polarization at the Fermi level and this fact can be exploited in a number of device applications. One way to measure such polarization is to employ spin-polarized voltage probes which could measure the spin-resolved voltages in these states subjected to a finite bias. In this work, we have explored two such possibilities for the surface states of quantum spin Hall system which is a two-dimensional topological insulator and also included disorder effects. Within our theoretical model, it is established that such measurements should be possible if the probes are designed appropriately~\cite{Tianelectrical2015} featuring robust signals even in the presence of disorder. 

\section{Acknowledgments}
VA acknowledges support from IISER Kolkata in the form of a subsistence grant. VA also wants to thank Rafiqul Rahaman for helping him with job submission in the cluster computing facility. KR thanks the sponsorship, in part, by the Swedish Research Council. SD would like to acknowledge the MATRICS grant (MTR/ 2019/001 043) from the Science and Engineering Research Board (SERB) for funding. We acknowledge the central computing facility (DIRAC supercomputer) and the computational facility at the Department of Physics (KEPLER) at IISER Kolkata.


\appendix

\section{Derivation of the scattering matrix elements at a junction of a chiral edge tunnel-coupled to the HES}\label{appA}

The scattering amplitudes for a wavefunction ${\psi}_\eta^{(j)}$ with $\eta\in\{R,L,R'\}$ across a point contact (PC) located at $x=x_j$ can be obtained by studying its equation of motion (e.o.m) 
\begin{equation}
 \imath \hbar\dot{\psi}_\eta^{(j)}=[\psi_\eta^{(j)},\mathcal{H}^{(j)}],
\end{equation}
where $\mathcal{H}^{(j)}=\mathcal{H}_{\rm HES}+\mathcal{H}_{\rm subprobe}^{(j)}+\mathcal{H}_{T}^{(j)}$ is specified in the main text (see Eq.~\ref{Hedge}-\ref{htun}). Integrating the e.o.m over a region from $-\epsilon$ and $\epsilon$ with the limit $\epsilon\rightarrow0$ across $x = x_j$, one obtains the required equations between the incoming and the outgoing amplitudes connected through the PC~\cite{wadhawan2018multi}. From now on, we drop the superscript $j$ assuming all of the following relations pertain to the $j$-th PC.

\begin{figure}
\centering
\includegraphics[width=1.0\columnwidth]{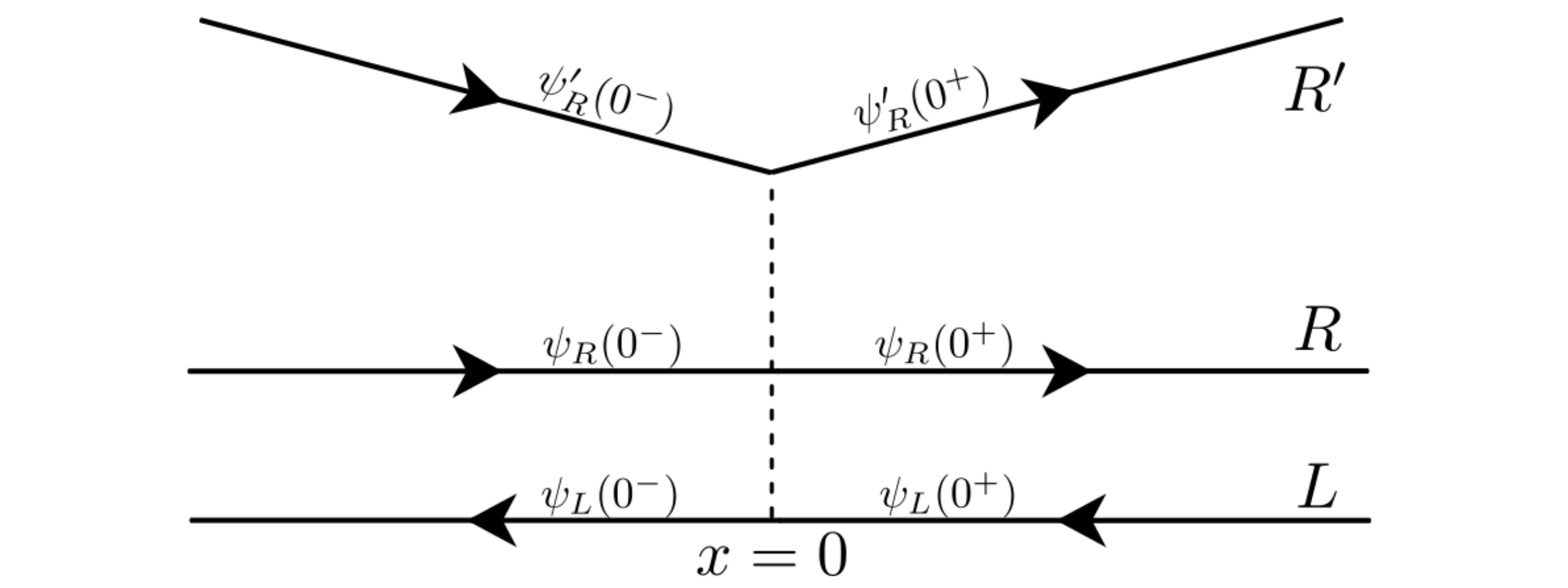}
\label{fig:figg11}
\end{figure}

The scattering matrix elements $s_{\eta \eta'}$, which appear in Eq.~(\ref{scat_mat}), are defined as 
\begin{equation}
\underbrace{\begin{pmatrix} \psi_R(0^+) \\ \psi_L(0^-) \\ \psi_{R'}(0^+) \end{pmatrix}}_{\psi^{\rm out}} = 
  \underbrace{\begin{pmatrix} s_{RR} & s_{RL} & s_{RR'} \\ s_{LR} & s_{LL} & s_{LR'} \\ s_{R'R} & s_{R'L} & s_{R'R'} \end{pmatrix}}_{S-{\rm matrix}} 
  \underbrace{\begin{pmatrix} \psi_R(0^-) \\ \psi_L(0^+) \\ \psi_{R'}(0^-) \end{pmatrix}}_{\psi^{\rm in}},
  \label{fullSmat}
\end{equation} 
where $s_{RR}=\psi_R(0^+)/\psi_{R}(0^-)$, $s_{LR}=\psi_{L}(0^-)/\psi_{R}(0^-)$, $s_{R'R}=\psi_{R'}(0^+)/\psi_{R}(0^-)$, and so on. 
The explicit expressions for these elements, after taking $v_F = 1$, read 
\begin{align}\label{bs1} 
s_{RR} &= [ 8+\iota (\Gamma_{RL}\Gamma_{RR^{'}}^{\ast}\Gamma_{LR^{'}}+\Gamma_{RL}^{\ast}\Gamma_{RR^{'}}\Gamma_{LR^{'}}^{\ast}) \nonumber \\
&~~~~~~-2\{\vert \Gamma_{RL}\vert^{2}+\vert\Gamma_{RR^{'}}\vert^{2}-\vert\Gamma_{LR^{'}}\vert^{2}\}]/{\cal D} \nonumber \\
s_{RL} &= \left[-8\iota \Gamma_{RL}-4\Gamma_{RR^{'}}\Gamma_{LR^{'}}^{\ast}\right]/{\cal D} \nonumber \\
s_{RR'} &= \left[-8\iota \Gamma_{RR^{'}}-4\Gamma_{RL}\Gamma_{LR^{'}}\right]/{\cal D} \nonumber \\
s_{LR} &= \left[-8\iota\Gamma_{RL}^{\ast}-4\Gamma_{RR^{'}}^{\ast}\Gamma_{LR^{'}}\right]/{\cal D} \nonumber \\
s_{LL} &= [8+\iota (\Gamma_{RL}\Gamma_{RR^{'}}^{\ast}\Gamma_{LR^{'}}+\Gamma_{RL}^{\ast}\Gamma_{RR^{'}}\Gamma_{LR^{'}}^{\ast}) \nonumber \\
&~~~~~~-2\{\vert \Gamma_{RL}\vert^{2}-\vert\Gamma_{RR^{'}}\vert^{2}+\vert\Gamma_{LR^{'}}\vert^{2}\}]/{\cal D} \nonumber \\
s_{LR'} &= \left[-8\iota \Gamma_{LR^{'}}-4\Gamma_{RL}^{\ast}\Gamma_{RR^{'}}\right]/{\cal D} \nonumber \\
s_{R'R} &= \left[-8\iota \Gamma_{RR^{'}}^{\ast}-4\Gamma_{RL}^{\ast}\Gamma_{LR^{'}}^{\ast}\right]/{\cal D} \nonumber \\
s_{R'L} &= \left[-8\iota \Gamma_{LR^{'}}^{\ast}-4\Gamma_{RL}\Gamma_{RR^{'}}^{\ast}\right]/{\cal D} \nonumber \\
s_{R^{'}R^{'}} &= [8+\iota (\Gamma_{RL}\Gamma_{RR^{'}}^{\ast}\Gamma_{LR^{'}}+\Gamma_{RL}^{\ast}\Gamma_{RR^{'}}\Gamma_{LR^{'}}^{\ast}) \nonumber \\
&~~~~~~-2\{-\vert \Gamma_{RL}\vert^{2}+\vert\Gamma_{RR^{'}}\vert^{2}+\vert\Gamma_{LR^{'}}\vert^{2}\}]/{\cal D},
\end{align} 
where the common denominator, ${\cal D}$ is 
\begin{equation}
\begin{split}
{\cal D}=8-\iota\left(\Gamma_{RL}\Gamma_{RR^{'}}^{\ast}\Gamma_{LR^{'}}+\Gamma_{RL}^{\ast}\Gamma_{RR^{'}}\Gamma_{LR^{'}}^{\ast}\right)
\\+2\{\vert \Gamma_{RL}\vert^{2}+\vert\Gamma_{RR^{'}}\vert^{2}+\vert\Gamma_{LR^{'}}\vert^{2}\}.
\end{split}
\end{equation}
The functional form of $\Gamma_{\eta\eta'}$ in terms of $t'$ and $\theta$ are noted in the main text from which all our results readily follow.

\section{Derivation of the scattering matrix elements for the FM barrier}\label{appB}
The $2 \times 2$ scattering matrix defined at the FM barrier of the Fig.~\ref{fig:figg4} can be easily obtained by following the procedure discussed in appendix~\ref{appA}. as 
\begin{equation}
\underbrace{\begin{pmatrix} \psi_R(x_0^+) \\\\ \psi_L(x_0^-)\end{pmatrix}}_{\psi^{\rm out}} = 
  \underbrace{\begin{pmatrix} s_{RR}^{FM} & s_{RL}^{FM} \\\\ s_{LR}^{FM} & s_{LL}^{FM} \end{pmatrix}}_{S-{\rm matrix}} 
  \underbrace{\begin{pmatrix} \psi_R(x_0^-) \\\\ \psi_L(x_0^+) \end{pmatrix}}_{\psi^{\rm in}}.
  \label{fullSmat2}
\end{equation} 
The explicit expressions for the $S$-matrix elements are given by
\begin{align}\label{bs3}
s_{RR}^{FM} &= \left[8  - 2 |\Gamma^{FM}_{RL}|^2\right]/{\cal D}^{\prime}\nonumber \\
s_{RL}^{FM} &=  \left[-8\iota \Gamma^{FM}_{RL}\right]/{\cal D}^{\prime}\nonumber \\
s_{LR}^{FM} &=  \left[-8\iota (\Gamma^{FM}_{RL})^*\right]/{\cal D}^{\prime}\nonumber \\
s_{LL}^{FM} &= \left[8  - 2 |\Gamma^{FM}_{RL}|^2\right]/{\cal D}^{\prime},
\end{align}
where the common denominator, ${\cal D}^{\prime}$ is
\begin{equation}
{\cal D}^{\prime} = \left[8  + 2 |\Gamma^{FM}_{RL}|^2\right].
\end{equation}
Similar functional forms of $\Gamma^{FM}_{\eta\eta'}$ follow with $t'$ being replaced by $\cal B$ and the angular dependence having no relevance. 

\bibliography{references}
\end{document}